\documentclass[twocolumn,nofootinbib,superscriptaddress]{revtex4-1}
\pdfoutput=1 

\usepackage{latexsym,epsfig,amssymb, amsmath,nicefrac}
\usepackage{color}
\usepackage[makeroom]{cancel}
  \usepackage{latexsym}
  \usepackage{epsf}
  \usepackage{amssymb}
  \usepackage{graphicx}
  \usepackage{amsmath}
  \usepackage{amsmath,amssymb,amsthm}
  \usepackage{verbatim}
    \usepackage{epstopdf}

\newcommand{\e}{\textrm{e}}

\def\tb0{\tilde{\beta}_0}
{\def\b0{\beta_0}

\def\bi{\begin{itemize}}
\def\ei{\end{itemize}}
\def\be{\begin{equation}}
\def\ee{\end{equation}}
\newcommand{\bea}{\begin{eqnarray}}
\newcommand{\eea}{\end{eqnarray}}
 
\renewcommand{\Im}{\textrm{Im}\,}
\renewcommand{\Re}{\textrm{Re}\,}

\def\Kahler{K\"{a}hler~}

\begin{document}

\begin{flushleft}
CERN-TH-2016-168 \\  DESY 16-143 \\  \textcolor{white}{blank space}
\end{flushleft}

\title{Moduli Backreaction on Inflationary Attractors}

\author{Diederik Roest,}
\email{d.roest@rug.nl}
\affiliation{Van Swinderen Institute for Particle Physics and Gravity, University of Groningen, \\ Nijenborgh 4, 9747 AG Groningen, The Netherlands}
\affiliation{Theoretical Physics Department, CERN, CH-1211 Geneva 23, Switzerland}
\author{Marco Scalisi}
\email{marco.scalisi@desy.de}
\affiliation{Van Swinderen Institute for Particle Physics and Gravity, University of Groningen, \\ Nijenborgh 4, 9747 AG Groningen, The Netherlands}
\affiliation{Deutsches Elektronen-Synchrotron DESY, Notkestra{\ss}e 85, 22607 Hamburg, Germany}
\author{Pelle Werkman}
\email{p.j.werkman@rug.nl}
\affiliation{Van Swinderen Institute for Particle Physics and Gravity, University of Groningen, \\ Nijenborgh 4, 9747 AG Groningen, The Netherlands}

\begin{abstract}

We investigate the interplay between moduli dynamics and inflation, focusing on the KKLT-scenario and cosmological  $\alpha$-attractors. General couplings between these sectors can induce a significant backreaction and potentially destroy the inflationary regime; however, we demonstrate that this generically does not happen for $\alpha$-attractors. Depending on the details of the superpotential, the volume modulus can either be stable during the entire inflationary trajectory, or become tachyonic at some point and act as a waterfall field, resulting in a sudden end of inflation. In the latter case there is a universal supersymmetric minimum where the scalars end up, preventing the decompactification scenario. The gravitino mass is independent from the inflationary  scale with no fine-tuning of the parameters. The observational predictions conform to the universal value of attractors, fully compatible with the Planck data, with possibly a capped number of e-folds due to the interplay with moduli.
\end{abstract}

\maketitle

%%%%%%%%%%%%%%%%%%%%%%%%%%%%%%%%%%%%%%%%%%%%%%%%%%%%%%%%%%%%%%%%%%%%%%

\section{Introduction}

Compactifications of string theory generically come with many \emph{moduli}: classically massless scalar fields that parametrize properties of the internal manifold and that give rise to unobserved long-range interactions.While one expects quantum effects to generate masses for these scalar fields, it is difficult to realize this while retaining computational control \cite{DineSeiberg}. In the case of Calabi-Yau compactifications, the moduli parametrize deformations of the manifold's \Kahler form, its complex structure and the string coupling. One may generate a mass for the latter two by turning on fluxes in the internal manifold \cite{GKP}. However, the \Kahler moduli cannot be stabilized in this manner. Instead, Kachru, Kallosh, Linde and Trivedi (KKLT) \cite{KKLT} argued that one can stabilize the \Kahler moduli using non-perturbative corrections while maintaining computational control.

The central issue we intend to address in this paper is how the presence of the moduli sector can affect an inflationary regime. Coupling inflation with other moduli generically leads to mutual \emph{backreaction}. On the one hand, the inflationary energy can destabilize the moduli. This was anticipated in \cite{Kallosh2004}, where it was shown that stabilizing the \Kahler modulus in the simplest model of inflation leads to a bound $H < m_{\nicefrac{3}{2}}$ on the inflationary Hubble scale $H$, related to the gravitino mass $m_{\nicefrac{3}{2}}$ in the vacuum after inflation\footnote{In the same paper \cite{Kallosh2004}, it was pointed out that using a specific combination of two exponentials in the superpotential generically improves the decoupling of the two physical scales. This so-called KL model and its coupling to inflation was further explored in \cite{Davis:2008fv,Kallosh:2011qk}.}. Conversely, the dynamics of the volume modulus may induce a backreaction which renders the inflaton scalar potential too steep to support inflation. 

The issue of moduli stabilization during inflation was subsequently investigated in an explicit string theory setup in \cite{Kachru:2003}. In this paper, a scalar field $r_1$ parametrizing the separation between an anti-brane and a brane serves as the inflaton. A warped geometry sourced by five-form fluxes generates a naturally flat potential for $r_1$. As described above, fluxes serve to stabilize all moduli except the volume modulus, which stabilize with a KKLT-like structure. The interplay of $r_1$ and the \Kahler modulus generically yields a large shift in the second slow-roll parameter $\eta$, thus spoiling inflation. 

More generally, the interplay between moduli stabilization and supersymmetry breaking has been extensively studied in literature (see e.g. \cite{Dudas:2012wi,Buchmuller:2014vda,Vercnocke}). For quadratic inflation, this topic has been investigated in detail at the supergravity level in \cite{Challenges}, where the super- and \Kahler potentials were sum separable between the moduli and inflaton sectors. In every setup considered in \cite{Challenges}, the naive stability bound $H < m_{\nicefrac{3}{2}}$ was verified and there was a destabilization of the \Kahler modulus on the inflationary trajectory, at large field values of the inflaton. Generating enough e-folds of inflation imposed stringent constraints on the parameter space. 

The aim of this paper is to study the interplay of KKLT-like moduli stabilization and supergravity $\alpha$-attractor models of inflation. The $\alpha$-attractor models provide an elegant description of the inflationary dynamics with robust predictions \cite{SuperconformalAttractors, Carrasco:2015uma,  Carrasco:2015rva} that are in excellent agreement with the latest data on the cosmic microwave background \cite{Planck2015,Ade:2015lrj,Ade:2015tva,Array:2015xqh}. Moreover, they have been coupled to various other sectors \cite{Kallosh:2015lwa,deSitterLandscape,Carrasco:2015pla,Kallosh:2016ndd,Kallosh:2016gqp,Kallosh:2016sej} (see \cite{Ueno:2016dim,Eshaghi:2016kne} for reheating constraints on this class of models). In this paper, we will investigate their resilience under moduli backreaction. Specifically, we will show that combining these two sectors together yields surprising consequences. While the \Kahler modulus will turn out to be always stable during inflation, the major effect of the backreaction can be instead beneficial. It generically induces/enhances the attractor inflationary regime as well as produces a supersymmetric vacuum where the scalars can sit at the end of their evolution. Remarkably, this allows to decouple the inflationary and SUSY breaking scales with no amount of fine-tuning.

We first review the supergravity descriptions of both $\alpha$-attractors and moduli stabilization in Sec.~\ref{SECtwosectors}, and outline the strategy of our analysis and the main physics traits arising from coupling these two moduli sectors.  We then proceed to discuss the vacuum structure and inflationary features of different coupling cases. In Sec.~\ref{Sec.Product}, we analyse the product coupling case while in Sec.~\ref{Sec.General} we show how the latter can be generalized with surprising physics outcomes. The resulting inflationary dynamics with concrete examples and predictions are the topics of Sec.~\ref{Sec.Dynamics}. In Sec.~\ref{Sec.GeneralNilpotent} we show how the corresponding construction simplifies in the presence of a nilpotent superfield. We conclude in Sec.~\ref{SECconclusions} with a summary of our results and future perspectives.

\vspace{-0.3cm}

\section{Review and strategy}\label{SECtwosectors}

\subsection{Inflation: $\alpha$-attractors}\label{SECalpha}

The inflationary class of models referred to as $\alpha$-attractors has its origin in an underlying superconformal or supergravity model \cite{SuperconformalAttractors, Carrasco:2015uma}. A crucial role in these theories is played by the \Kahler manifold spanned by the scalars: it is taken to be a hyperbolic manifold of maximal symmetry and negative curvature. For a wide range of superpotentials of such theories, the resulting inflationary scenarios share a common trait: inflation takes place around the boundary of the field space leading to exponential fall-off terms from a de Sitter phase. 

While the first constructions of $\alpha$-attractors conform to the general supergravity structure of \cite{Kawasaki, Rube} and have an inflaton and a stabilizer superfield, it turns out that a more minimal set-up is possible involving only the inflaton superfield \cite{AlphaScale,Linde:2015uga} (alternatively, one can also employ a vector supermultiplet \cite{Ferrara:2013rsa}). We will mainly focus on the chiral single-superfield formulation of \cite{AlphaScale}, and comment on the extension with a second superfield in Sec.~\ref{Sec.GeneralNilpotent}. The minimal theory consists of a \Kahler and superpotential given by\footnote{Alternatively, one could work in a \Kahler frame related by the transformation
\begin{align}
K_{\alpha} &\rightarrow K_{\alpha} - \frac{3\alpha}{2}\log(\Phi) - \frac{3\alpha}{2}\log(\bar{\Phi}) = -3\alpha \log\left(\frac{\Phi + \bar{\Phi}}{|\Phi|}\right) \,,\\
W_{\alpha} &\rightarrow W_{\alpha}\Phi^{-\frac{3\alpha}{2}}, \label{shift-K}
\end{align}
that makes the rescaling and inversion symmetries of the \Kahler potential explicit \cite{Carrasco:2015uma}. The rescaling symmetry translates into a shift symmetry of the canonically normalized inflaton.}
 \begin{align}\label{KWalpha}
  K_{\alpha} = -3 \alpha \log(\Phi + \bar \Phi) \,, \quad  W_{\alpha} = \Phi^{n_-} - \Phi^{n_+} f(\Phi) \,,
 \end{align}
defined in terms of the monomial powers
 \begin{align}
  n_\pm =\tfrac{3}{2}\big(\alpha \pm \sqrt{\alpha}\big) \,.
 \end{align}
The \Kahler geometry is maximally symmetric and has constant curvature given by $R_K = - 2 / (3 \alpha)$. 
 
The superpotential contains two monomial factors, which are related to each other by a \Kahler transformation and an inversion of $\Phi$. Taken separately, these factors yield a Minkowski vacuum along the trajectory in field space defined by $\Phi = \bar{\Phi}$ (i.e. $\Im(\Phi) = 0$). This generalizes the no-scale structure \cite{Cremmer:1983bf,Ellis:1983sf} that appears for $\alpha =1$ and has been dubbed {\it $\alpha$-scale} model \cite{AlphaScale} as the curvature $R_K$ determines stabilization of the imaginary direction. When both powers are included in the superpotential, the cross term between them generates a non-vanishing cosmological constant, whose sign is opposite to the relative sign between the two factors. When $\alpha > 1$, the imaginary direction of $\Phi$ is stabilized with a high mass at $\Im(\Phi) = 0$, so that a truncation to $\Phi = \bar{\Phi}$ is consistent.

Finally, a function $f(\Phi)$ is included to introduce an \emph{inflationary profile} instead of a flat de Sitter line. If this function is expressible as a generic Taylor expansion around $\Phi = 0$, and it induces a negative linear term, then inflation will occur\footnote{If $f(\Phi)$ multiplies the lower power $\Phi^{n_-}$ in Eq.~\eqref{KWalpha}, then the physics remains unchanged when we expand the system around $\Phi \rightarrow \infty$.}. Along $\Phi=\bar{\Phi}$ and when expressed in terms of the canonically normalized field 
\begin{equation} \label{changevariable}
\varphi = -\sqrt{\tfrac{3\alpha}{2}}\ \log(\Phi)\,,
\end{equation}
the potential around the singular point $\Phi = 0$ is an exponentially suppressed deviation from a flat plateau
\begin{equation}\label{POTattractors}
V = V_0 - V_1 \e^{-\sqrt{\frac{2}{3\alpha}}\varphi} + \ldots.
\end{equation} 
on which inflation can occur. The resulting observables are very simple and predictive: at lowest order in $1/N$, the inverse of the number of e-folds, these read
 \begin{align}
 n_s = 1 - \frac2N \,, \quad r = \frac{12 \alpha}{N^2} \,, \label{usual-predictions}
 \end{align}
which agrees very well with the latest cosmological data \cite{Planck2015,Ade:2015lrj,Ade:2015tva,Array:2015xqh} for an $\alpha$ of order unity\footnote{Formally, the expression \eqref{usual-predictions} agrees with the data for $\alpha \lesssim 30$, but the approximations used to derive \eqref{usual-predictions} break down earlier. For $\alpha \gg 1$, $\alpha$-attractors tend to converge to monomial models such as $V \sim \Phi^n$ (see e.g.\cite{SuperconformalAttractors,DoubleAttractors,AlphaScale})}.

\subsection{Volume stabilization: KKLT}

The second model is the KKLT scenario for volume modulus stabilization \cite{KKLT}. At the supergravity level, it simply consists of the following \Kahler- and superpotential:
 \begin{align}\label{KWKKLT}
   K_{\rm mod} = -3 \log(T + \bar T) \,, \quad W_{\rm mod} = W_0 + A \exp( - a T) \,.
 \end{align}
The logarithmic, no-scale like \Kahler potential is naturally generated by flux compactification (see \cite{Grana:2005jc,Douglas:2006es} for some reviews on this topic). The constant $W_0$ is meant to contain the contribution of other moduli, which are assumed to be stabilized supersymmetrically at a very high scale by some other mechanism \cite{GKP}. The exponential term is a non-perturbative correction, generated by gluino condensation or Euclidean $D3$-branes.
 
The background in which this setup is embedded is an orientifolded Calabi-Yau flux compactification of the type discussed above. KKLT assume a Calabi-Yau with cohomology $h^{(1,1)} = 1$, so that there is only a single complex \Kahler modulus.

The scalar potential generated by these $K_{\rm mod} $ and $W_{\rm mod}$ has an AdS minimum at $T=T_0$ defined by
\begin{equation}\label{eq:SUSYKKLT}
3 W_0 = - e^{-aT_0}(3 + 2aT_0) A \,,
\end{equation}
which is also the condition for unbroken supersymmetry, $D_T W = 0$. To stabilize $T$ at a large positive value, as is necessary for the consistency of the approximation scheme, one must choose $W_0$ smaller than and opposite in sign to $A$. Furthermore, the potential has the Dine-Seiberg Minkowski minimum at $T \rightarrow \infty$, which corresponds to the decompactification of the internal space. This run-away minimum is always present in the scalar potential. 

\subsection{Coupling and backreaction}

The main aim of this paper is to investigate the backreaction of moduli on the inflationary dynamics. In particular, we will show that $\alpha$-attractors suffer from negligible backreaction of the moduli sector in many cases. This special immunity is mainly due to the fact that inflation happens at the boundary of moduli space ($\Phi\rightarrow0$ or $\Phi\rightarrow\infty$). Here the original KKLT AdS minimum can be lifted to dS thanks to the $\alpha$-scale mechanism, while the supersymmetric properties of the $T$ sector remain unchanged.  Then, the field $\Phi$ can drive inflation while rolling down along a stable minimum defined by $D_T W = 0$. Upon switching to the canonical variable $\varphi$, the stretching of the boundary to a long plateau provides the key to understanding why $\delta T$ is minimized during the inflationary evolution (an analogous behaviour has been noticed in \cite{Kallosh:2016gqp}, where the interaction between $\alpha$-attractors and matter fields is exponentially suppressed). Thus the backreaction is negligible during the expansion period. Note that this behaviour is the opposite of what was noticed in \cite{Challenges} for the quadratic inflationary scenario. 

Towards the end of inflation, the situation turns out to be non-trivial as it usually depends on the specific couplings in the superpotential. Generically, the mass of the modulus $T$ becomes lighter and it can produce a destabilization point. However, we will show that a number of interesting things happen in the case of $\alpha$-attractors and KKLT: First of all, one can always avoid the destabilization point by means of a specific profile function $f$. Secondly and more importantly, we will prove the existence of a stable {\it universal vacuum} at finite value of $T$ which can prevent the inflaton to run away towards the decompactification limit $T\rightarrow\infty$. We will explain all these results in detail in the following sections.

Throughout this paper we will consider the case of additive \Kahler potentials,
 \begin{align}\label{Kahlercombined}
 K = K_{\alpha} + K_{\rm mod} \,,
 \end{align}
in the combined model of inflation and moduli stabilization. For the superpotential, we will instead consider a variety of combinations which maintain the $\alpha$-scale properties at the boundary of the moduli space.

\section{Product coupling}\label{Sec.Product}
We start by considering a superpotential which is product separable in the moduli and inflaton sectors:
\begin{equation}\label{eq:productsuper}
W = W_{\rm mod}(T)\ W_{\alpha}(\Phi)\,,
\end{equation} 
where the factors are given by Eq.~\eqref{KWalpha} and Eq.~\eqref{KWKKLT}. This corresponds to a sum separable \Kahler function $G=K + \ln(W) + \ln(\bar W)$, which is defined when $W \neq 0$. The product coupling reduces the mixing between both sectors, as found for hybrid inflation \cite{PostmaDavis}. Moreover, it allows for important simplifications; e.g.~the supersymmetric critical points of the $T$ or $\Phi$ sectors remain so in the combined theory \cite{Achucarro,Achucarro:2008sy}.

In the following, we separately analyse the cases without and with an inflationary profile in $W_{\alpha}$. In the first case ($f=1$), we present the non-trivial vacuum structure which arises from the interplay of the two sectors. In the latter case ($f=f(\Phi)$), we show how a consistent inflationary dynamics can be implemented in this context.

\subsection{$\alpha$-KKLT and universal Minkowski vacuum}

\begin{figure}[t!]
	\vspace{0.4cm}
	\begin{center}
		\includegraphics[width=8.5cm]{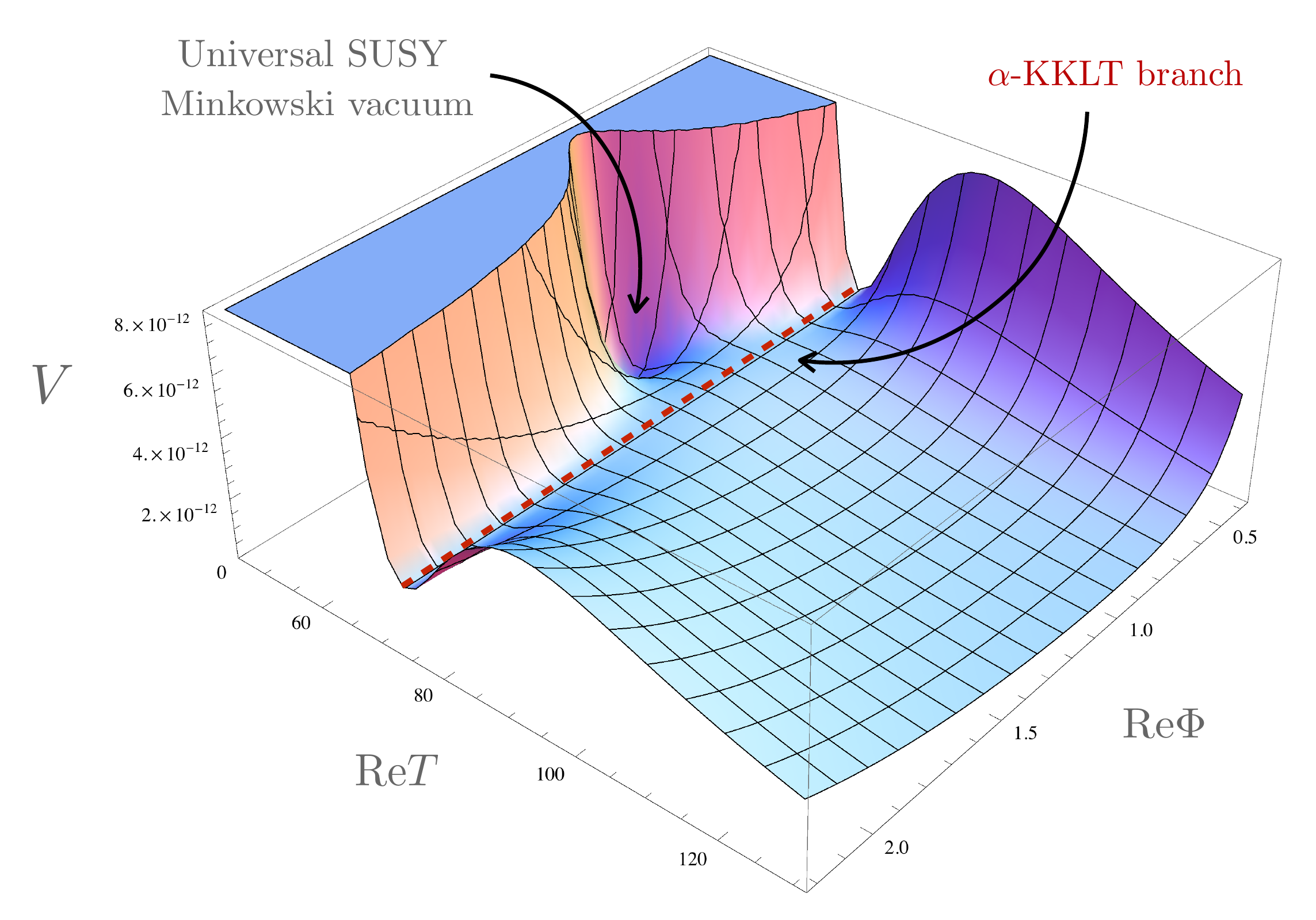}
			\vspace{0.cm}
		\caption{\it Plot of the scalar potential defined by Eq.~\eqref{eq:productsuper} as function of $\Re(T)$ and $\Re(\Phi)$. The $\alpha$-KKLT branch, placed at $T_0$ with potential value given by Eq.~\eqref{dSalphaKKLT}, is highlighted by the red dashed line. It develops an instability around $\Phi=1$ where one can clearly appreciate the universal Minkowki minimum, placed at shifted values of $T$. (Parameters: $A = 1$, $W_0 = -0.0004$, $a = 0.1$, $\alpha = 1.5$.)}
		\label{fig:valley}
	\end{center}
	\vspace{-0.5cm}
\end{figure}

Let us take $f=1$ and consider the following superpotential:
\begin{equation}\label{eq:productsuper}
W(\Phi, T) = \left[\Phi^{n_-} - \Phi^{n_+}\right]\left[W_0 + A \exp( - a T)\right]\,.
\end{equation} 
The product structure of $W$ allows to easily uplift the original KKLT minimum by means of the same $\alpha$-scale mechanism described in Sec.~\ref{SECalpha}. For this reason, we will refer to this branch as {\it $\alpha$-KKLT}. It is characterized  $D_T W=0$ and represents an extremum of the full model at $T=T_0$, given by Eq.~\eqref{eq:SUSYKKLT}, independently of the value of the field $\Phi$ (see Fig.~\ref{fig:valley}). Along this branch, supersymmetry in the $\Phi$-direction is spontaneously broken yielding a positive dS phase with
\begin{align}\label{dSalphaKKLT}
V = \frac{2^{1-3 \alpha } a^2 A^2 e^{-2 a T_0}}{3 T_0}\,.
\end{align} 
Although the product separable superpotential guarantees a critical point along this branch, it does not imply stability. In general, this trajectory may be instead a maximum or an inflection point in the $T$-direction.

In the left and right asymptotic limits ($\Phi \rightarrow 0$ and $\Phi \rightarrow \infty$, respectively), $T = T_0$ is a minimum in the $\Re(T)$ direction. To see this, note that the scalar potential defined by Eq.~\eqref{eq:productsuper} has terms proportional to $\Phi^{-3\sqrt{\alpha}}$ and $\Phi^{3\sqrt{\alpha}}$, which dominate in the left and right asymptotic limits, respectively. These terms are non-negative and cancel exactly when the supersymmetry condition Eq.~\eqref{eq:SUSYKKLT} is substituted. Therefore, the $\alpha$-KKLT trajectory $T = T_0$ is a minimum at the asymptotic limits; the $\Phi^{\pm 3\sqrt{\alpha}}$ terms generate a large mass for $\Re(T)$ in the left and right asymptotic limits, which is an opposite behaviour to what was found in \cite{Challenges}.

\begin{figure}[htb]
	\begin{center}
		\includegraphics[width=7.9cm]{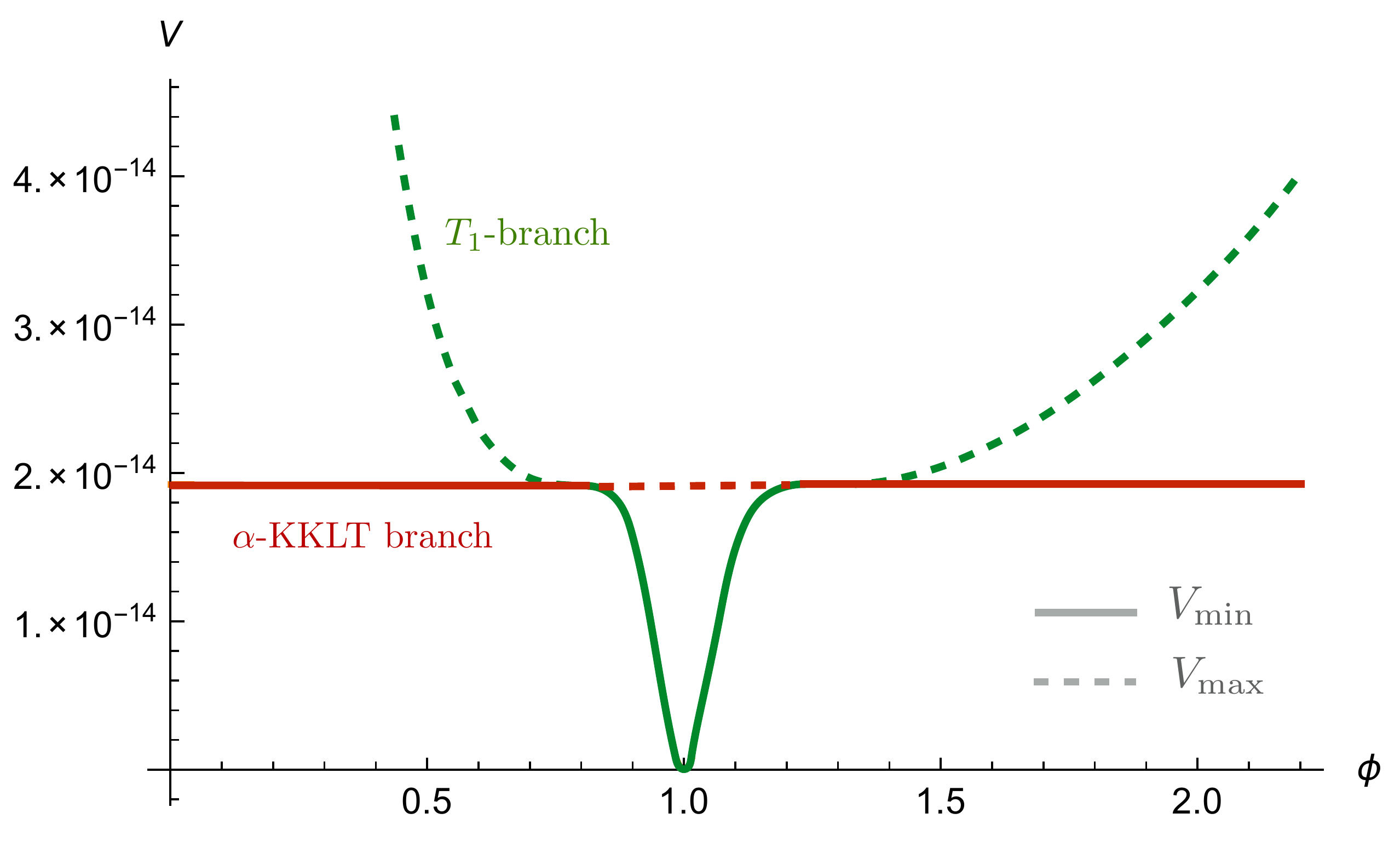}\vspace{0.1cm}

			\includegraphics[width=7.2cm]{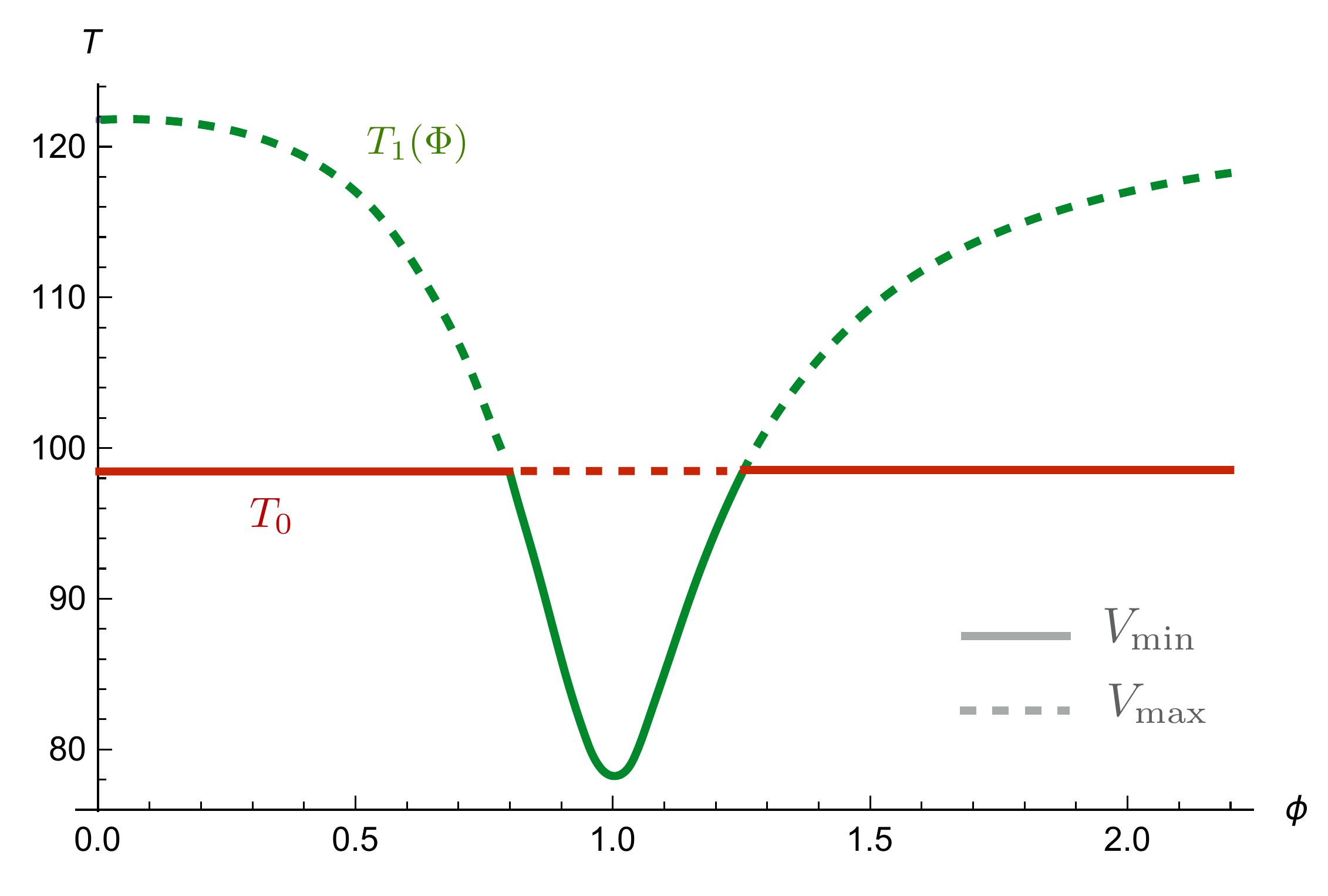}
		\caption{\it Effective potential $V(\Phi)$ (upper panel) and vacuum structure (lower panel) for the product separable case. The minimum of the potential in $T$ is denoted by the solid lines, the maximum with the dashed ones. The red line denotes the $\alpha$-KKLT branch at $T_0$ while the green line represents the non-SUSY branch defined by Eq.~\eqref{eq:SecondBranch}. (Parameters: $A = 1$, $W_0 = -0.0004$, $a = 0.1$, $\alpha = 1.1$)}
		\label{fig:scalarproduct}
	\end{center}
	\vspace{-0.5cm}
\end{figure}

In the intermediate region between the asymptotic limits $\Phi \rightarrow 0$ and $\Phi \rightarrow \infty$, the vacuum structure is different (see Fig.~\ref{fig:valley}). In particular, there is a {\it universal Minkowski vacuum} that has both product factors of the superpotential vanishing, rendering it supersymmetric. When $W_{mod}(T) = 0$, (i.e. at $T = T_S = - {\log(-W_0/A)}/a$), the scalar potential reads:
\begin{equation}
V(\Phi, T_S)=\frac{2^{-3 \alpha -1} a^2 A^2 e^{-2 a T_S} \Phi ^{-3 \sqrt{\alpha }} \left(\Phi ^{3 \sqrt{\alpha }}-1\right)^2}{3 T_S}\,.
\end{equation}
At $\Phi = 1$, we find  a supersymmetric Minkowski minimum (regardless of the choice of parameters). This minimum is part of a second branch of solutions to $\partial_T V = 0$, distinct from the $\alpha$-KKLT trajectory. This branch of solutions corresponds to solutions $T=T_1(\Phi)$ of the following equation:
\begin{align}\label{eq:SecondBranch}
& -9\left(1 + \Phi^{3\sqrt{\alpha}}\right)^2e^{aT_1} W_0 = \bigg(9 + 8aT_1 + 4a^2 T_1^2 + 9\Phi^{6 \sqrt{\alpha}}  \notag \\
& - 16 a T_1 \Phi^{3\sqrt{\alpha}} +18\Phi^{3\sqrt{\alpha}} + 8 a T_1 \Phi^{6 \sqrt{\alpha}} + 4a^2 T_1^2 \Phi^{6\sqrt{\alpha}}\bigg) A \,,
\end{align}
which is obtained by first solving $\partial_T V = 0$ for $W_0$. The latter approach lets us clearly differentiate different branches of solutions to $\partial_T V = 0$, even though it may exclude some of them. At $\Phi = 1$, Eq.~\eqref{eq:SecondBranch} reduces to $W_0 = -Ae^{-aT_1}$, to which the solution is $T_1 = T_S$. Therefore, this branch of solutions includes the universal SUSY Minkowski vacuum. At the asymptotic limits, Eq.~\eqref{eq:SecondBranch} determines the location of the maximum in between the $\alpha$-KKLT branch and the Dine-Seiberg Minkowski minimum at $T \rightarrow \infty$. Fig.~\ref{fig:scalarproduct} illustrates this situation.

The extremal branch $T_1(\Phi)$ defined by Eq.~\eqref{eq:SecondBranch} intersects the $\alpha$-KKLT branch in two points in field space. These are inflection points in the $T$-direction and the \Kahler modulus becomes massless. After the first inflection point, the $\alpha$-KKLT branch becomes a maximum and the other branch a minimum, until the second inflection point on the other side of \mbox{$\Phi = 1$}. The locations $\Phi_\pm$ of the inflection points are obtained by equating the two expressions Eq.~\eqref{eq:SUSYKKLT} and Eq.~\eqref{eq:SecondBranch}:
\begin{equation}
\left(1 + 2aT_0\right) \Phi_\pm^{3 \sqrt{\alpha}} = 7 + 2 a T_0 \pm \frac{2\sqrt{6(2 + aT_0)}}{1 + 2 aT_0} \,,
\end{equation}
where $T_0$ is the solution to Eq.~\eqref{eq:SUSYKKLT}.

\subsection{Inserting an inflationary profile}\label{ProdInflationaryProfile}
One can tilt the original flat positive plateau of the $\alpha$-KKLT branch and produce a consistent inflationary regime by turning the profile function $f(\Phi)$ on in $W$. The superpotential of the full model then reads 
\begin{equation}
W(\Phi, T) = \left[\Phi^{n_+}f(\Phi) - \Phi^{n_-}\right]\left[W_0 + Ae^{-aT}\right]\,.
\end{equation}
This procedure is  analogous to the single-superfield case presented in \cite{AlphaScale} and summarized in Sec.~\ref{SECalpha}. Along the $\alpha$-KKLT branch, the scalar potential becomes
\begin{equation}
V = \frac{a^2 A^2 \left[\Phi  f'(\Phi )+3 \sqrt{\alpha } f(\Phi )\right] \left[3 \sqrt{\alpha }+\Phi ^{3 \sqrt{\alpha }+1} f'(\Phi )\right]}{2^{3 \alpha -1}\ 27 \alpha  T_0\ e^{2 a T_0}} \,,
\end{equation}
which is identical to the potential generated by Eq.~\eqref{KWalpha},  up to a $\Phi$-independent rescaling. For $f$ being a generic expansion in the geometric field $\Phi$, the scalar potential at large values of the canonical inflaton $\varphi$, becomes an exponential deviation from dS, as given by Eq.~\eqref{POTattractors}. 

The stability conditions of the real directions and vacuum structure features of the model with an inflationary slope $f(\Phi)$ strikingly resemble the ones described above with $f=1$. In addition to the tilted $\alpha$-KKLT which is still supersymmetric in the $T$ sector, there is generically a second branch which breaks supersymmetry in both directions. Moreover, for a large class of choices of $f$, there is always a supersymmetric Minkowski minimum along the latter. To see this, we substitute $T = T_S$ in the scalar potential to obtain:
\begin{equation}
V(\Phi)=\frac{2^{-3 \alpha -1} a^2 A^2 \Phi ^{-3 \sqrt{\alpha }} e^{-2 a T_S} \left(\Phi ^{3 \sqrt{\alpha }} f(\Phi )-1\right)^2}{3 T_S} \,.
\end{equation}
The universal Minkowski vacuum is therefore located at $\Phi = \Phi_S = f(\Phi_S)^{-1/3\sqrt{\alpha}}$, when the latter gives a real solution for $\Phi$.

At the asymptotic boundaries of the moduli space (very small/large values of $\Phi$), the $\alpha$-KKLT extrema are stable minima, serving as a perfect starting point for inflation.
At finite values of $\Phi$, depending on the choice of $f$, the two branches may intersect and produce again two inflection points. However, now these yield important consequences for the inflationary dynamics: The \Kahler modulus becomes indeed massless, which implies that the single-field effective description breaks down. Running off to the Dine-Seiberg decompactification limit is one of the greatest risks. Nevertheless, this is not the only option for the inflaton field, which can instead follow a safe route and still produce a consistent cosmological scenario with a proper vacuum. Two possible scenarios arise indeed at this point:
\begin{itemize}
\item An appropriate choice of the profile function $f(\Phi)$ may lead to the end of inflation and produce a phenomenologically suitable vacuum before the inflection point. This completely  avoids  the possibility of decompactification of the internal manifold, as the \Kahler modulus never becomes massless. In this case, inflation proceeds as it does in the single-field $\alpha$-scale model \cite{AlphaScale}. The observable predictions are the usual Eq.~\eqref{usual-predictions}.
\item Generic and non-fine-tuned choices of $f(\Phi)$ will have instead the inflaton coming across one of the two inflection points. The subsequent dynamical evolution of the scalar fields is difficult to anticipate and usually depends on the initial conditions. It can happen that the \Kahler modulus $T$ runs off to the Dine-Seiberg decompactification limit. Alternatively, it may follow the second branch $T_1$ of minima towards the universal Minkowski vacuum. Here, all directions become stabilized again at a high mass scale and the scalar trajectories are quite predictable. In this scenario, inflation ends immediately at the inflection point as $\Re(T)$ experiences a large shift which renders the effective potential too steep. This is a \emph{waterfall effect} akin to that in models of hybrid inflation \cite{Linde:1993cn}. It introduces a positive shift $\Delta N$ in the effective number of e-folds probed by inflation, as the abrupt end moves the inflationary window further up on the scalar potential plateau. The predictions thus become:

 \begin{align}\label{predictionsdeltaN}
 n_s = 1 - \frac{2}{N+\Delta N} \,, \qquad r = \frac{12 \alpha}{(N + \Delta N)^2} \,.
 \end{align} 
\end{itemize}

The above qualitative descriptions of the two possible situations will be complemented by concrete examples with dynamical simulations in Sec.~\ref{Sec.Dynamics}.

\section{General coupling}\label{Sec.General}

We can generalize the product separable model, presented above, in a way that preserves the concave inflationary valley represented by the stable $\alpha$-KKLT branch at the boundary. We then consider the following superpotential: 
\begin{align}\label{eq:GeneralCouplingSuper}
W &= \Phi^{n_-} W_-(T) - \Phi^{n_+} W_+(T) \,,
\end{align}
where we allow for two distinct KKLT structures
 \begin{align}
  W_{\pm} = W_0 + A_\pm \exp(-a T) \,.
 \end{align}
Note that the constant parts can be set equal by rescaling $\Phi$, while for simplicity we have chosen to keep the power in the non-perturbative term identical (which would be natural if they follow from the same non-perturbative physics). The deviation from the product separable case is therefore parametrized by the ratio $A_+ / A_-$.

\subsection{$\alpha$-KKLT at the boundary}

In the general coupling case, when $A_+ \neq A_-$, the original $\alpha$-KKLT branch ($D_T W = 0$) is no longer guaranteed to be a critical point. Furthermore, its location $T_0$ is no longer a constant, but becomes a function of $\Phi$. However, in the left and right asymptotic limits, the situation turns out to be analogous to the product case and the fluctuation of $T_0$ becomes negligible. At the boundary, the $\alpha$-KKLT structure remains indeed unperturbed: it has unbroken supersymmetry in the $T$ direction and a stable de Sitter minimum. The intuitive reason is because, at the asymptotic limits, one of the $\Phi^{n_\pm}$ terms dominates over the other and $W$ can be effectively considered again product separable. We explicitly show this below.

The equation that determines $T_0$ ($D_T W|_{T = T_0, \Phi} = 0$) as a function of $\Phi$ is
\begin{equation}\label{eq:SUSYCondition}
W_+'(T) - \Phi^{-3\sqrt{\alpha}}W_-'(T)  =  \frac{3}{2T}\left[W_+(T) - \Phi^{-3 \sqrt{\alpha}}W_-(T)\right]\,,
\end{equation}
which is symmetric under the simultaneous interchange $\Phi^{-3\sqrt{\alpha}} \rightarrow \Phi^{3\sqrt{\alpha}}$, $W_+(T) \rightarrow W_-(T)$. As $\Phi \rightarrow 0$, the latter equation reduces to  the KKLT condition Eq.~\eqref{eq:SUSYKKLT},
\begin{equation}\label{PhizeroDWT}
W_-'(T) = \frac{3}{2T}W_-(T)\,.
\end{equation} 
Therefore, in the left asymptotic limit, the piece $W_-$ alone determines the $D_T W = 0$ asymptotic trajectory. The analogous statement holds in the $\Phi \rightarrow \infty$ limit with $W_+$. 

Furthermore, one can calculate the scalar potential by means of Eq.~\eqref{Kahlercombined} and Eq.~\eqref{eq:GeneralCouplingSuper} and substitute the full SUSY condition for $T$ Eq.~\eqref{eq:SUSYCondition}. One then finds
\begin{equation}\label{eq:GeneralCouplingPlateau}
V = \frac{3\cdot\ 2^{-3 \alpha -1} W_-(T_0) W_+(T_0)}{T_0^3}\,,
\end{equation}
where $T_0$ is implicitly a function of $\Phi$. This is de Sitter at the asymptotic limits as long as $W_-(T)$ and $W_+(T)$ have the same sign at each other's supersymmetric points. Therefore, along the $D_T W =0$ trajectory, the situation is just analogous to the single superfield case of \cite{AlphaScale} as well as to the product case of Sec.~\ref{Sec.Product}, along the $\alpha$-KKLT trajectory.

The stability of this branch is guaranteed at the boundary. The full scalar potential contains a term 
\begin{equation}
\frac{2^{-3 (\alpha +1)} \Phi ^{-3 \sqrt{\alpha }} \left[3 W_-(T)-2 T W_-'(T)\right]^2}{3 T^3}\,,
\end{equation}
which dominates in the limit $\Phi \rightarrow 0$. It vanishes just when the SUSY condition for the $T$ sector Eq.~\eqref{PhizeroDWT} is substituted. Since the term is non-negative, this asymptotic $\alpha$-KKLT trajectory turns out to be a stable minimum in the $T$ direction, as we found in the product separable case.  The \Kahler modulus $T$ becomes then highly massive on the inflationary plateau. This is again an inversion of the behaviour highlighted in \cite{Challenges}, where the $T$ becomes light (and later tachyonic) in the region of field space where inflation happens.

\subsection{Universal inflation from backreaction}

Moving away from the boundary, the shift in $T_0$ generates a small fall-off from de Sitter as a backreaction. This effect is the main qualitative difference between the product separable case and this more general model. The backreaction renders the effective potential suitable for slow-roll inflation even without introducing a non-trivial profile function $f(\Phi)$. We show this below.

The shifts of the position $T_0$ are $\mathcal{O}(\Phi^{3\sqrt{\alpha}})$ in the $\Phi \rightarrow 0$ limit, and $\mathcal{O}(\Phi^{-3\sqrt{\alpha}})$ in the opposite $\Phi \rightarrow \infty$ limit.  Near the left asymptotic limit, the first fall-off from de Sitter is then of the form
\be
V = V_0 - V_1\ \Phi^{3\sqrt{\alpha}} +\ldots\,.
\ee
It is interesting to notice that the power of $\Phi$ is not generic but specifically depends on $\sqrt{\alpha}$. This exponent cancels the $\alpha^{-1/2}$ in the relation \eqref{changevariable} between the geometrical inflaton $\Phi$ and the canonically normalized field $\varphi$, and we  have
\begin{equation}
\Phi^{3\sqrt{\alpha}} = e^{-\sqrt{\frac{18}{3}}\varphi}\,.
\end{equation}
The parameter $\alpha$ therefore drops out of the predictions for $n_s$ and $r$ in this model:
\begin{align}\label{PredictionsGeneralCoupling}
 n_s = 1 - \frac{2}{N} \,, \quad r = \frac{4}{3N^2} \,,
\end{align}
at leading order in the inverse of the number of e-folds $N$ and for any order unity $\alpha$. Note that, unlike the usual $\alpha$-attractors' predictions Eq.~\eqref{usual-predictions}, this scenario provides a precise value for the expected amount of primordial gravitational waves, independently of the value of the \Kahler curvature. 

It seems remarkable that the simple coupling of two moduli sectors naturally induces an inflationary regime with universal cosmological predictions given by Eq.~\eqref{PredictionsGeneralCoupling}, without inserting any profile function\footnote{One may consider the role of the profile function $f(\Phi)$ as parametrizing freedom in the model. However, in the light of building concrete string theory models, its presence might become an obstacle due to the difficulty of generating higher powers than cubic in $\Phi$.} $f(\Phi)$. Yet more intriguing is the strong insensitivity of the inflationary predictions to the details of model. Eq.~\eqref{PredictionsGeneralCoupling} would indeed hold for a very large class of parameter choices. We will provide a concrete and detailed investigation of the inflationary dynamics of this model in Sec.~\ref{Sec.Dynamics}.

\subsection{Vacuum structure and universal AdS minimum}

\begin{figure}[t]
	\vspace{0.4cm}	
	\begin{center}
		\includegraphics[width=7.7cm]{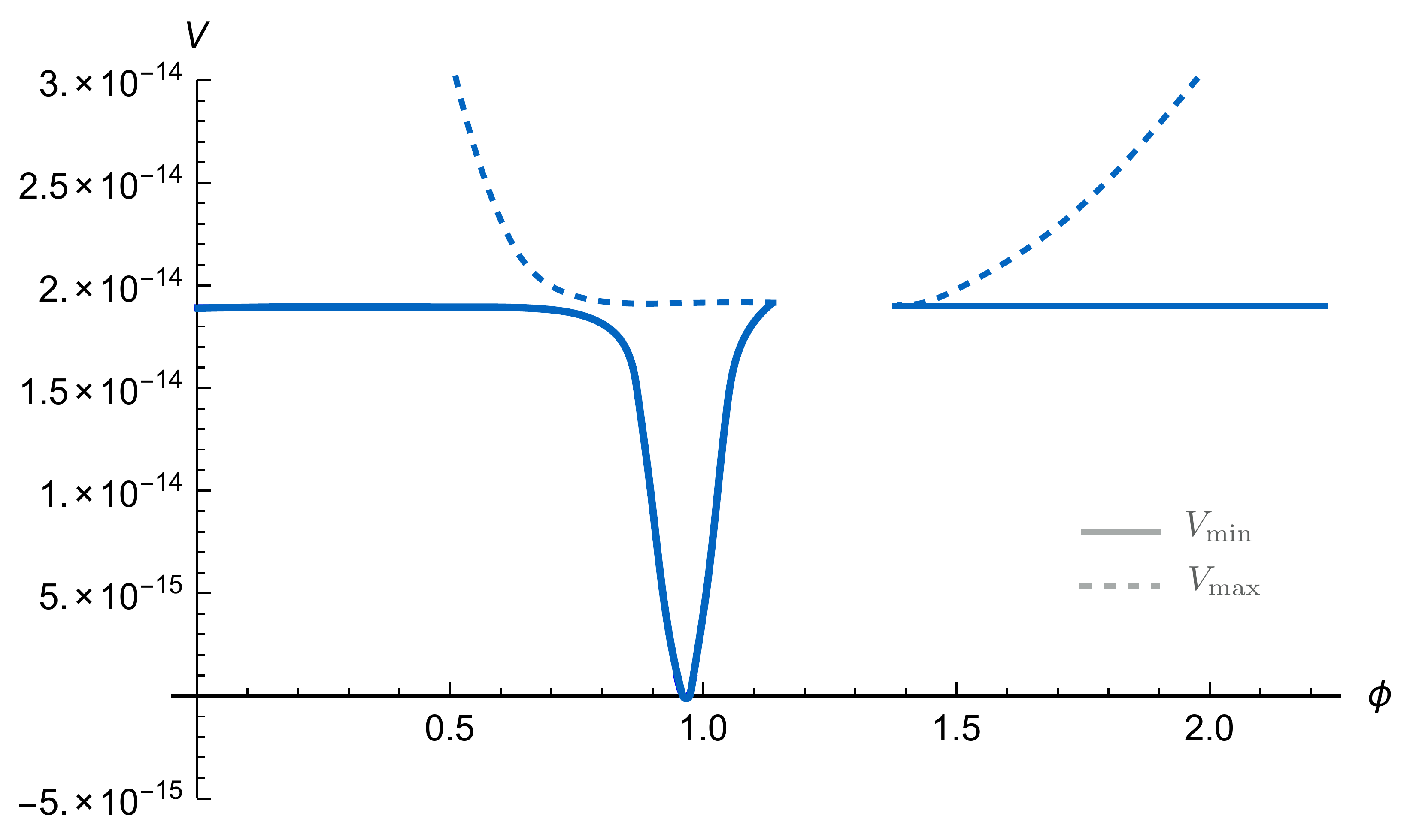}
		\hspace{0.2cm}
			\includegraphics[width=7.3cm]{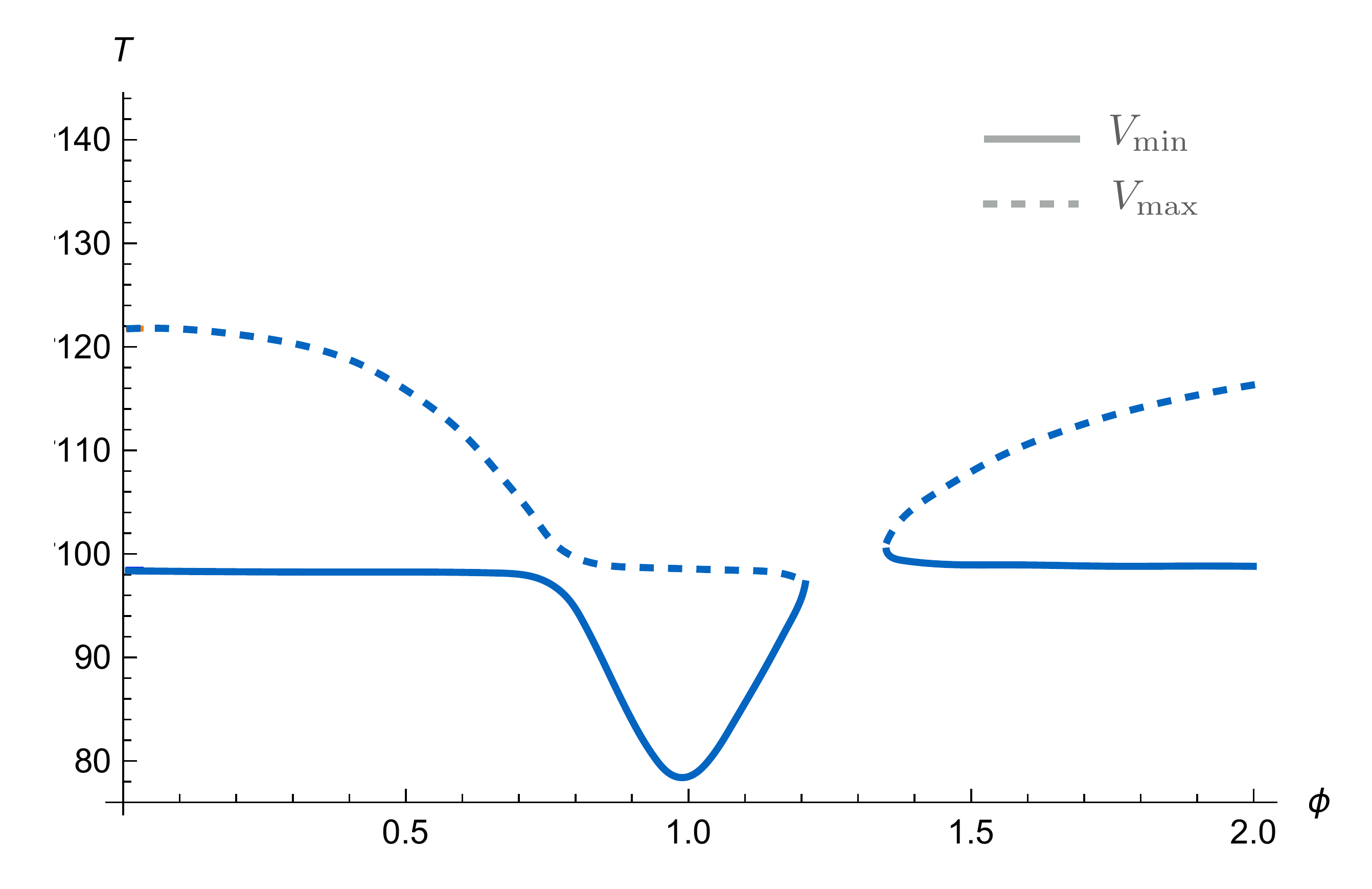}
		\caption{\it Effective potential $V(\Phi)$ (upper panel) and vacuum structure (lower panel) for the general coupling case. The minimum of the potential in $T$ is denoted by the solid line, the maximum by the dashed line. (Parameters: $A_+ =1.1$, $A_- = 1$, $W_0 = -0.0004$, $a = 0.1$ and $\alpha = 1.1$).}
		\label{fig:scalargeneral}
	\end{center}
	\vspace{0cm}
\end{figure}

In the intermediate region of the field space, the interplay of both KKLT terms $W_\pm$ is very important. The vacuum structure is markedly different from the product separable case. Specifically, the structure of the inflection points at finite $\Phi$ changes drastically when we allow for $A_+ \neq A_-$. We will discuss this as a continuous deformation from the product separable model. An infinitesimal change with $A_+ > A_-$ has the following effect on the two inflection points, as illustrated in Fig.~\ref{fig:scalargeneral}:
 \begin{itemize}
\item
The left inflection point disappears: the two extremal branches (that correspond to the $\alpha$-KKLT and the non-supersymmetric trajectories in the $\Phi \rightarrow 0$ limit) no longer meet and rather are located at different values of $T$, along the entire range of $\Phi$ up to the second inflection point. This is good news as it implies that the \Kahler modulus never becomes massless along the inflationary trajectory, depicted in dark blue in Fig.~\ref{fig:scalargeneral}. However, there is still a sudden shift in the \Kahler modulus which ends inflation in a waterfall effect. We will explore this in detail in Sec.~\ref{Sec.Dynamics}.
\item
The right one splits up in two new inflection points at closeby values in $\Phi$; in between these two inflection points, there are no critical points in the scalar potential. This is indicated by the empty regions in $\Phi$ of Fig.~\ref{fig:scalargeneral}, which implies that $T$ would run off to the asymptotic Dine-Seiberg minimum.
\end{itemize}

The corresponding effective potential is pictured on the left panel of Fig.~\ref{fig:scalargeneral}. 
This also shows the asymptotic $\Phi^{-3\sqrt{\alpha}}$ fall-off from de Sitter due to backreaction induced by the \Kahler modulus (note however that the inflationary potential is not expressed in terms of the canonical field $\varphi$). Upon making the difference between both moduli functions more pronounced, the vacuum structure is further deformed along these lines. For $A_+ < A_-$, the situation is mirror reversed.

The last noteworthy difference between product separable and general $\alpha$-scale coupling concerns the universal SUSY vacuum. We find this minimum at $(\Phi_S,T_S)$ once we impose $\partial_T V=\partial_{\Phi} V=0$ which returns the following equations: 
 \begin{align}
T_S =& \frac{1}{a}\log \left[\frac{A_+ \Phi_S^{3\sqrt{\alpha}}+A_-}{-W_0(\Phi_S^{3\sqrt{\alpha}}+1)}\right], \notag \\ 
3 \Phi_S ^{3 \sqrt{\alpha }} (A_--A_+) =&\left(\Phi_S ^{3 \sqrt{\alpha }}+1\right) \left(A_+ \Phi_S ^{3 \sqrt{\alpha}}- A_-\right)\notag \\
 &\times\log \left[\frac{A_+ \Phi_S^{3 \sqrt{\alpha }}+A_-}{-W_0(\Phi_S ^{3 \sqrt{\alpha }}+1)}\right]\,.
 \end{align}
This minimum is connected continuously to the asymptotic $\alpha$-KKLT trajectory, as can be seen from Fig.~\ref{fig:scalargeneral}. The minimum is induced by a large shift in the \Kahler modulus, which creates a waterfall effect that ends inflation immediately. Furthermore, it is supersymmetric as $D_T W = D_\Phi W = 0$. However, as the superpotential does not vanish it is an AdS vacuum. This is in line with the general theorem that infinitesimal changes to the theory will deform a SUSY Minkowski vacuum to a SUSY AdS one \cite{Vercnocke} (see \cite{Linde:2014ela} for a practical application).

\section{Inflationary dynamics}\label{Sec.Dynamics}

\subsection{Product coupling with inflationary profile}

As outlined above in Sec.~\ref{ProdInflationaryProfile}, in the product separable case with inflationary profile $f(\Phi)$, there are two distinct scenarios for inflation which both end in an appropriate Minkowski minimum. 

\begin{figure}[htb]	
	\begin{center}
		\includegraphics[width=7.2cm]{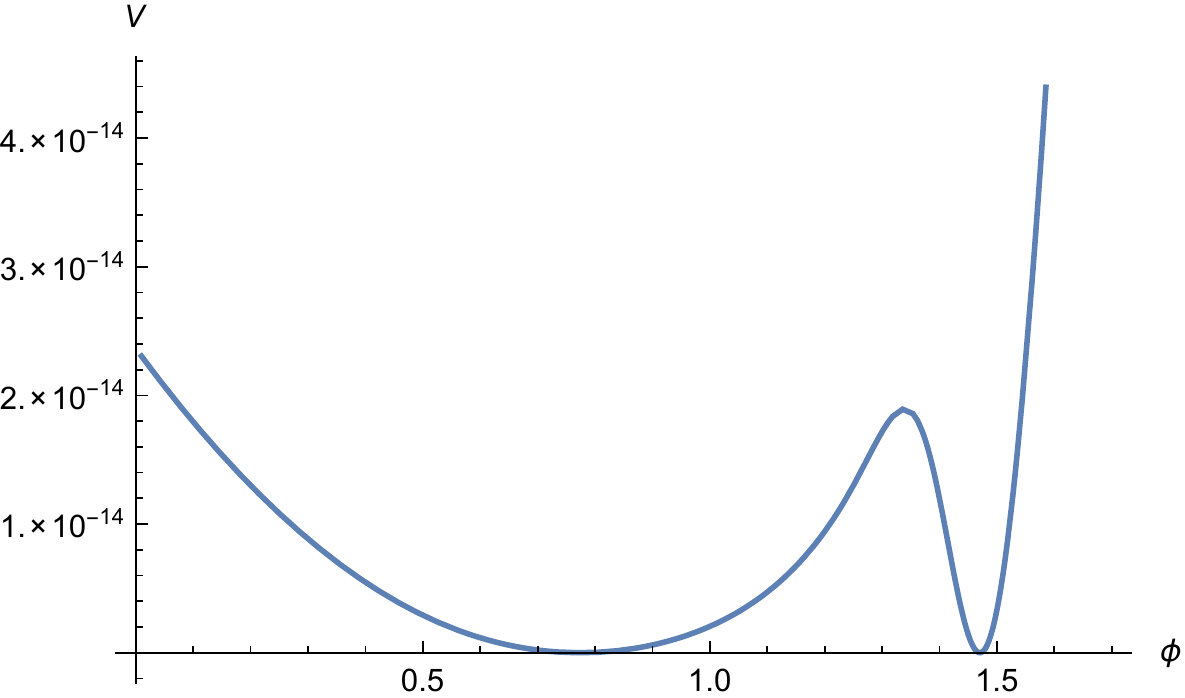}\vspace{0.2cm}
		\includegraphics[width=7.2cm]{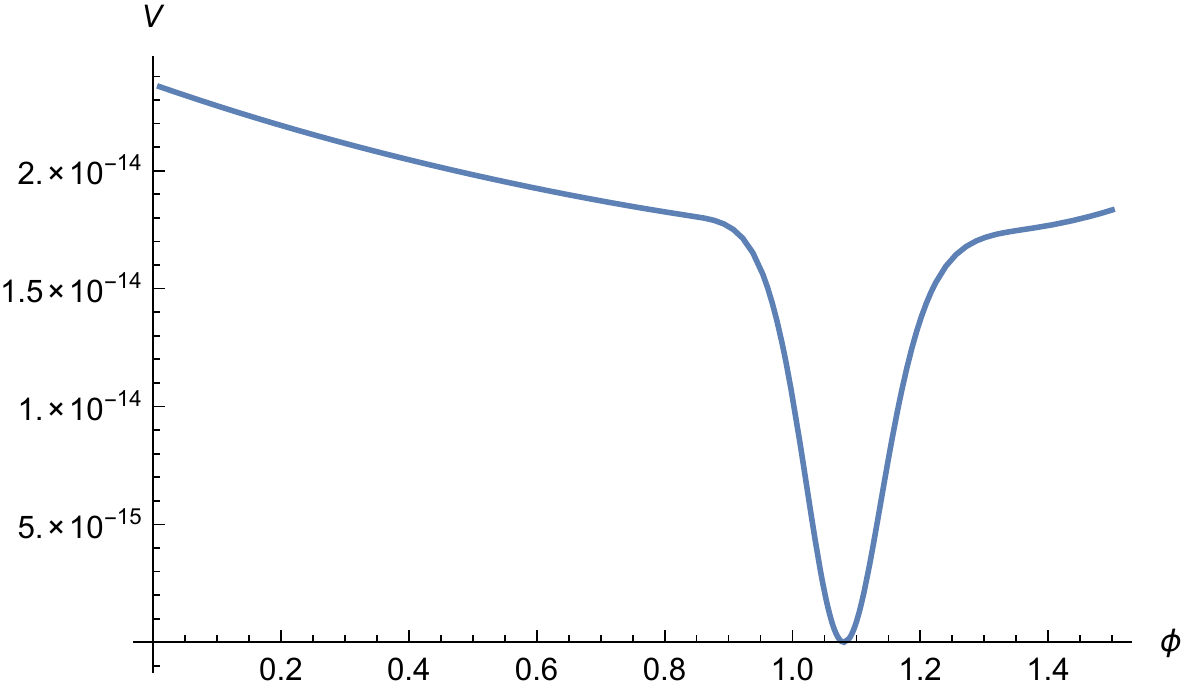}
		\caption{\it The effective scalar potential for product separable models with tuned profile \eqref{tuned} (upper panel) and generic profile \eqref{generic} (lower panel). Note the presence of the universal Minkowski minimum in both cases. In the tuned setup, there is an additional Minkowski minimum before the waterfall point. (Parameters: $W_0 = -0.0004$, $A = 1$, $a = 0.1$, $\alpha = 1$.)}
		\label{fig:quadraticperturbation}
	\end{center}
	\vspace{0cm}
\end{figure}

In the first, one simply tunes the profile function $f(\Phi)$ to produce a minimum along the $D_T W = 0$ trajectory, before the $T$-direction inflection point. The profile function $f(\Phi)$ must be of at least quadratic order in $\Phi$ for this purpose. Therefore we will consider 
 \begin{align} 
  f(\Phi) = c_0 + c_1 \Phi + c_2\Phi^2 \,, \label{tuned}
 \end{align}
with the following parameter values:
\begin{equation}
	c_0=1\,, \quad c_1 = -\frac{2 \sqrt{3}\sqrt{(2 \sqrt{\alpha} + 3\alpha)c_0 c_2}}{\lvert 1 - 9\alpha\rvert \lvert 1 -3\sqrt{\alpha}\rvert^{-1}}\,,\quad c_2=1\,.
\end{equation}
This tuning on the parameters  generates a Minkowski minimum, whose location is determined by the choice of $c_0$ and $c_2$. The resulting scalar potential is given in the left panel of Fig.~\ref{fig:quadraticperturbation} and has two Minkowski minima. The first of these occurs on the asymptotic $\alpha$-KKLT trajectory and is generated by the fine-tuned inflationary profile function. The second is the universal Minkowski minimum on the non-SUSY branch, which has been shifted to $\Phi_S$ as given by $\Phi_S = f(\Phi_S)^{-1/3\sqrt{\alpha}}$.

The maximum in between the two Minkowski minima occurs at the $T$-direction inflection point. To the left of the maximum, the \Kahler modulus is entirely stable, becoming very highly massive on the inflationary plateau close to $\Phi = 0$. Any deviation in $T$ from the supersymmetric value $T_0$ is quickly suppressed by the high mass on the inflationary plateau. The inflaton then proceeds to the $\alpha$-KKLT branch Minkowski minimum, where it first oscillates and then settles during reheating. In this scenario, there is essentially no dynamics in $T$ and the observable predictions are \eqref{usual-predictions}.

In the second scenario, we leave $f(\Phi)$ generic and make use of the universal Minkowski minimum along the non-SUSY trajectory. An example is given by
 \begin{align}
 f(\Phi) = 1 - 0.3 \Phi + 0.1 \Phi^2 \,, \label{generic}
 \end{align}
whose effective scalar potential is plotted in the right panel of Fig.~\ref{fig:quadraticperturbation}. In the case of product separable superpotential, this scenario can be dangerous because the \Kahler modulus becomes massless at the inflection point along the inflationary trajectory. It is necessary that $T$ does not run off to infinity after the inflection point is reached. However, the scenario is interesting because the inflaton settles into a minimum which is SUSY Minkowski without any fine-tuning. 

\begin{figure}[htb]
	\vspace{0.cm}	
	\begin{center}
		\includegraphics[width=7cm]{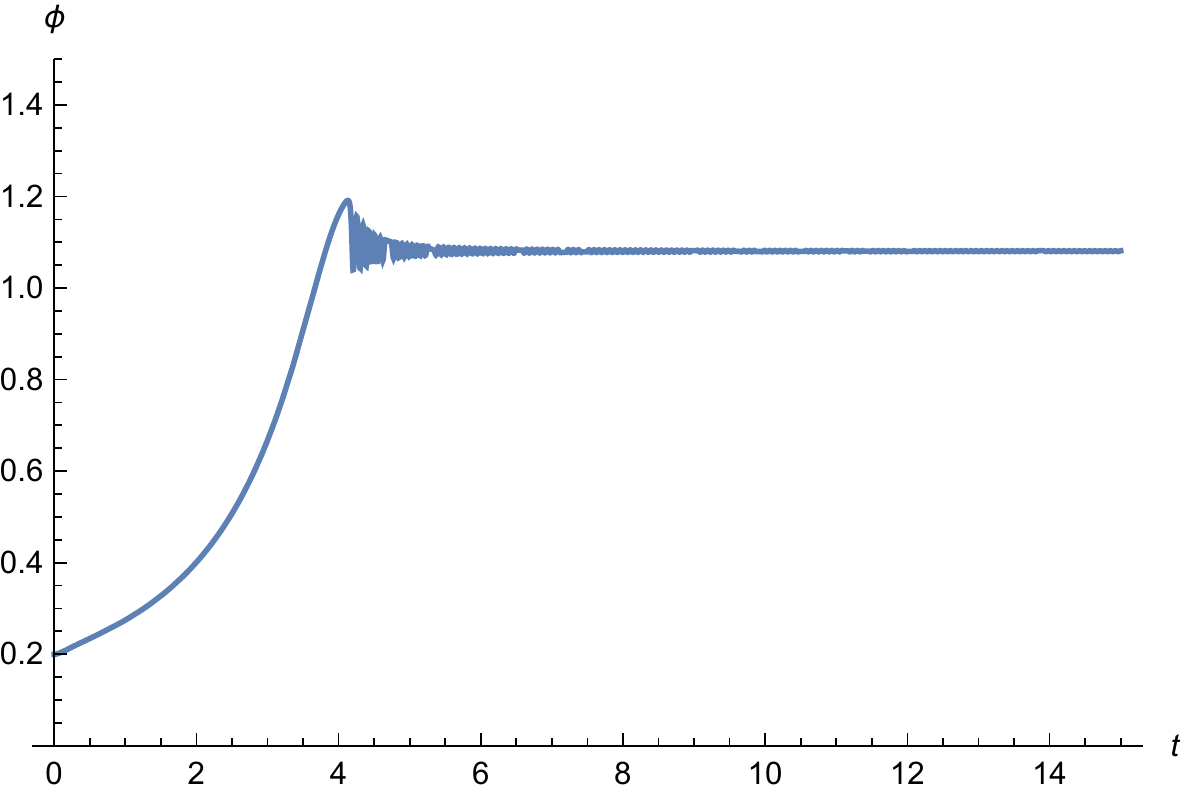}\vspace{0.3cm}
		\includegraphics[width=7cm]{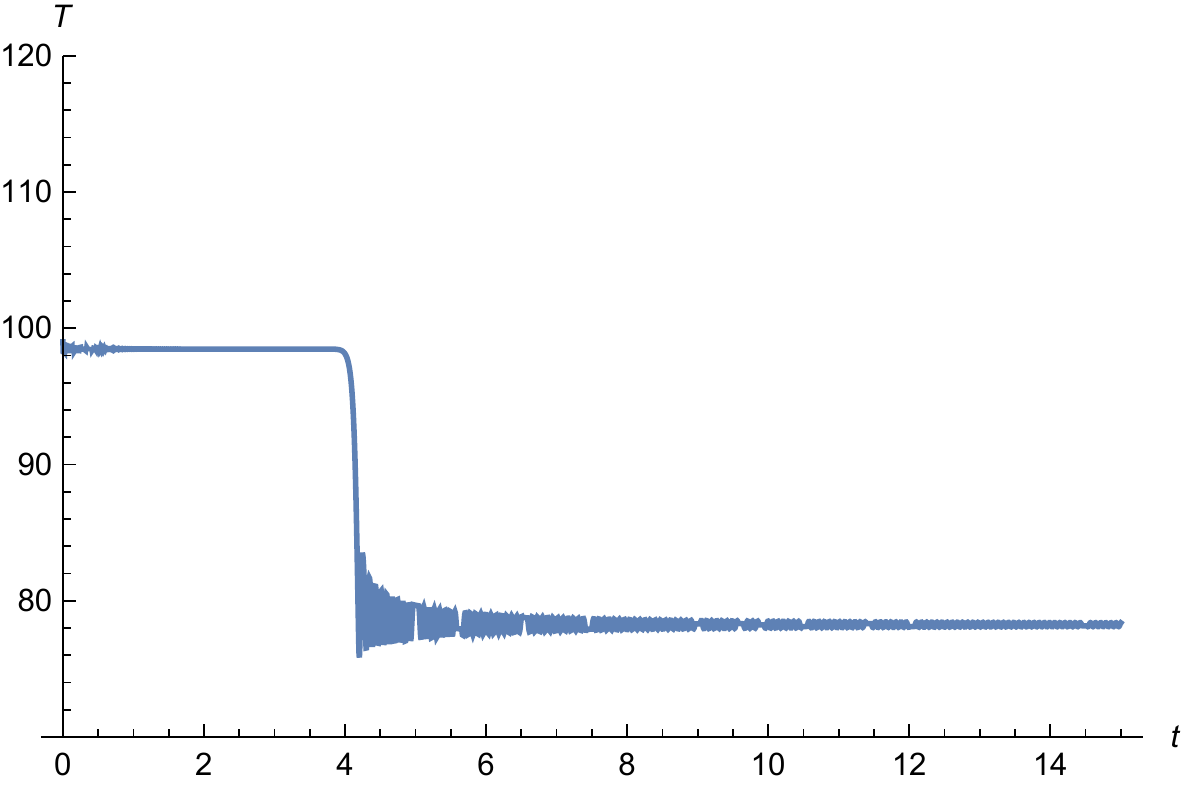}
		\caption{\it Simulation of scalar field dynamics of $\Phi$ (upper panel) and $T$ (lower panel) versus time for the product coupling case with generic profile \eqref{generic}. Initial displacements in $T$ from the $\alpha$-KKLT trajectory are quickly dampened close to $\Phi = 0$. The waterfall effect is clearly visible in this picture: the scalars oscillate around the universal Minkowski minimum after producing inflation on the plateau.  (Parameters:  $W_0 = -0.0004$, $A = 1$, $a = 0.1$, {$\alpha = 1$})}
		\label{fig:smallshiftsimulation}
	\end{center}
	\vspace{0cm}
\end{figure}

Figure \ref{fig:smallshiftsimulation} shows the result of a simulation of the scalar field dynamics in this scenario. With these initial conditions ($T_{i} = 99$, {$\Phi_i = 0.2$}), the scalars evolve towards the Minkowski minimum along the non-SUSY trajectory. When the inflection point is reached, a waterfall happens which ends inflation almost immediately. This induces a shift $\Delta N$ in the effective number of e-folds which determines the observable predictions Eq.~\eqref{predictionsdeltaN}. With this choice of parameters, this shift is small, $\Delta N \simeq 1$, as the inflection point occurs at a point in $\Phi$-space which is close to where inflation would have ended had $T$ been fixed at its supersymmetric value. The small shift keeps the predictions firmly in the observationally favored region. 

In this example, we end up at a Minkowski minimum without fine-tuning (see Fig.~\ref{fig:quadraticperturbation}) and generate viable observable predictions. There are two caveats: firstly, with a different choice of initial conditions, the scalars can evolve to the decompactification limit instead. We cannot calculate the probability of avoiding the decompactification, because there is no sensible definition for a prior distribution in initial conditions. However, if we imagine that inflation starts higher up on the plateau than pictured here, then the high mass scale of the real \Kahler modulus should dampen any deviations from $T = T_0$ quickly. This implies that the chosen initial conditions are quite sensible. Secondly, with a different $f(\Phi)$, the shift in the effective number of e-folds $\Delta N$ can push $n_s$ too close to unity.

The choice $\alpha = 1$ requires some additional considerations due to the stability of the imaginary directions, as we will discuss in a moment. However, the results on the effective scalar potentials do not change qualitatively when we change $\alpha$ by a small amount to e.g.~$\alpha=1.1$.

\subsubsection*{Stability of the imaginary directions}
We have examined the scalar potential along the trajectory defined by $\Phi = \bar{\Phi}$ and $T = \bar{T}$, i.e. at $\Im(\Phi) = \Im(T) = 0$. In order for this truncation to be consistent, the imaginary directions of both complex scalars must have a positive mass of at least order Hubble scale along this trajectory. In the case of product separable coupling, the scalar potential along the supersymmetric trajectory $T = T_0$ is equivalent to the single-superfield $\alpha$-scale model in the whole complex $\Phi$ space, up to a rescaling. It follows that the stability analysis of \cite{AlphaScale} carries over to our model. For $\alpha > 1$, the imaginary $\Phi$ direction is positive and divergent in the limit $\Phi \rightarrow 0$. For $\alpha < 1$, it is negative and divergent, and for $\alpha = 1$ it is negative and has a finite limit as $\Phi \rightarrow 0$. The most important contributions to the mass are:
\begin{equation}
M^2_{\Im(\Phi)} = \frac{a^2 (\alpha -1) A^2 e^{-2 a T} \Phi^{-3 \sqrt{\alpha }}}{2^{3 \alpha-1 }\ 9 \alpha  T} -\frac{a^2 A^2 c_0 e^{-2 a T}}{9 T}.
\end{equation}
When $\alpha \neq 1$, the $\Phi^{-3\sqrt{\alpha}}$ term dominates in the inflationary limit. When $\alpha = 1$, the negative constant term determines the mass.

\begin{figure}
	\includegraphics[width=7.2cm]{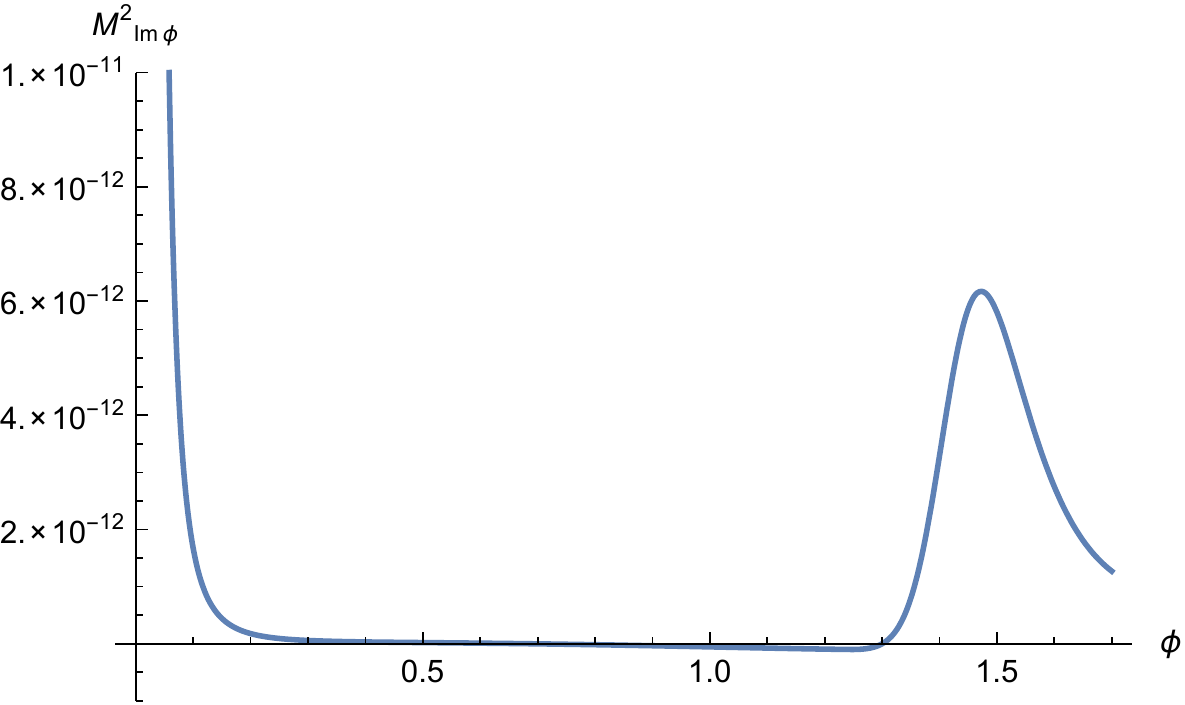}\vspace{0.3cm}
		\includegraphics[width=7.2cm]{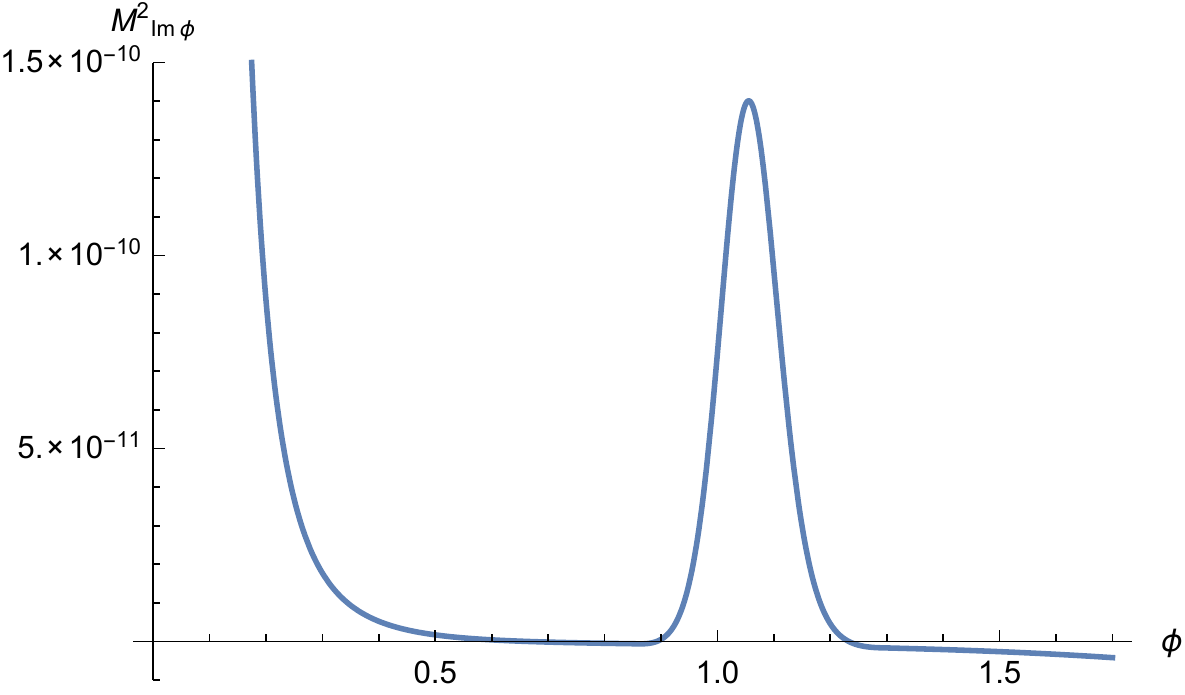}
	\caption{\it Mass of $\Im(\Phi)$ for the product case examples with tuned profile \eqref{tuned} (upper panel) and generic profile \eqref{generic} (lower panel) for $\alpha > 1$. In both cases, there is a divergent positive mass at the boundary $\Phi = 0$, a large positive mass around the SUSY Minkowski vacuum and a mild destabilization at intermediate field values.   (Parameters: $W_0 = -0.0004$, $A = 1$, $a = 0.1$)}
	\label{fig:ImaginaryMassProduct}
\end{figure}

At the universal SUSY Minkowski minimum all directions are stable and become very highly massive. The expression for the mass of $\Im(\Phi)$ at the vacuum reads:
\begin{equation}
M^2_{\Im(\Phi)} = \frac{a^2 A^2 e^{-2 a T} \Phi_S^{-3 \sqrt{\alpha }} \left[3 \sqrt{\alpha }+\Phi_S^{3 \sqrt{\alpha }+1}
	f'(\Phi_S)\right]^2}{2^{3 \alpha-1 }\ 9 \alpha  T}
\end{equation}
where $\Phi_S = F(\Phi_S)^{3\sqrt{\alpha}}$. Clearly, this mass is universally non-negative. The same can be shown for the other directions in the SUSY Minkowski vacuum.

In the intermediate region, the stability of $\Im(\Phi)$ is dependent on the choice of inflationary profile function $f(\Phi)$. For a generic choice of $f(\Phi)$, there is some instability at intermediate field values which flattens out (but does not disappear) as $\alpha$ is increased (see Fig.~\ref{fig:ImaginaryMassProduct}). There are several ways to deal with this apparent issue. Firstly, the high mass of $\Im(\Phi)$ in the inflationary limit (when $\alpha > 1$) dampens fluctuations away from $\Im(\Phi) = 0$ quickly. Furthermore, the instability occurs close in canonical variables to the $\Re(T)$ waterfall which ends inflation. These effects conspire to make the $\Im(\Phi)$ instability generally irrelevant for the classical inflationary dynamics. The scalar fields reach the completely stable region around the SUSY Minkowski vacuum before $\Im(\Phi)$ fluctuates enough to upset inflation. Once again, we find a simplification of the dynamics due to the asymptotic freedom of the inflaton.

Secondly, one can use the methods described in \cite{HyperbolicGeometry} to stabilize $\Im(\Phi)$ universally for any $\alpha$. The inflaton sector \Kahler potential $K_\alpha(\Phi, \bar{\Phi})$ can be generalized in the following way:
\begin{align}
K_{\alpha}(\Phi, \bar{\Phi}) =& - \frac{3\alpha}{1 + 2k_2}\log\bigg[\frac{\Phi + \bar{\Phi}}{|\Phi|}\bigg(1 + k_2 \frac{(\Phi - \bar{\Phi})^2}{(\Phi +  \bar{\Phi})^2}\notag \\ 
&+ k_4 \frac{(\Phi - \bar{\Phi})^4}{(\Phi +  \bar{\Phi})^4}  + \ldots\bigg)\bigg],
\end{align}
in the shift-symmetric \Kahler frame \eqref{shift-K}. The ellipses stand for higher even powers of $\frac{(\Phi - \bar{\Phi})}{(\Phi +  \bar{\Phi})}$, which preserve the inversion and dilatation invariance of the \Kahler potential. Taking $k_2, \ k_4, \ \ldots \rightarrow 0$, we recover the usual shift-symmetric \Kahler potential. The higher-order terms have no effect on the scalar potential along $\Phi = \bar{\Phi}$, but they dramatically alter the mass of the imaginary directions. A suitable choice of parameters renders $\Im(\Phi)$ stable over the entire field range. 

\subsection{General coupling without inflationary profile}

\begin{figure}[htb]
	\begin{center}
		\includegraphics[width=7.2cm]{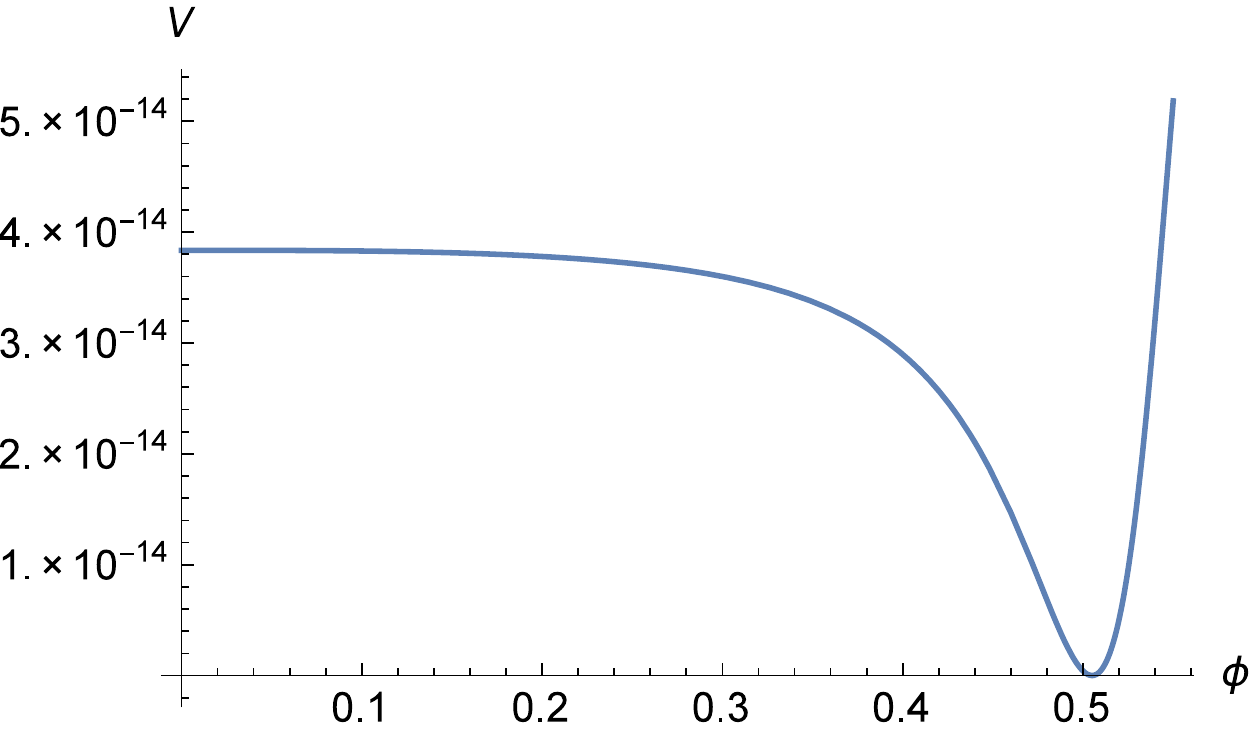}	\vspace{0.3cm}	
		
		\includegraphics[width=7.2cm]{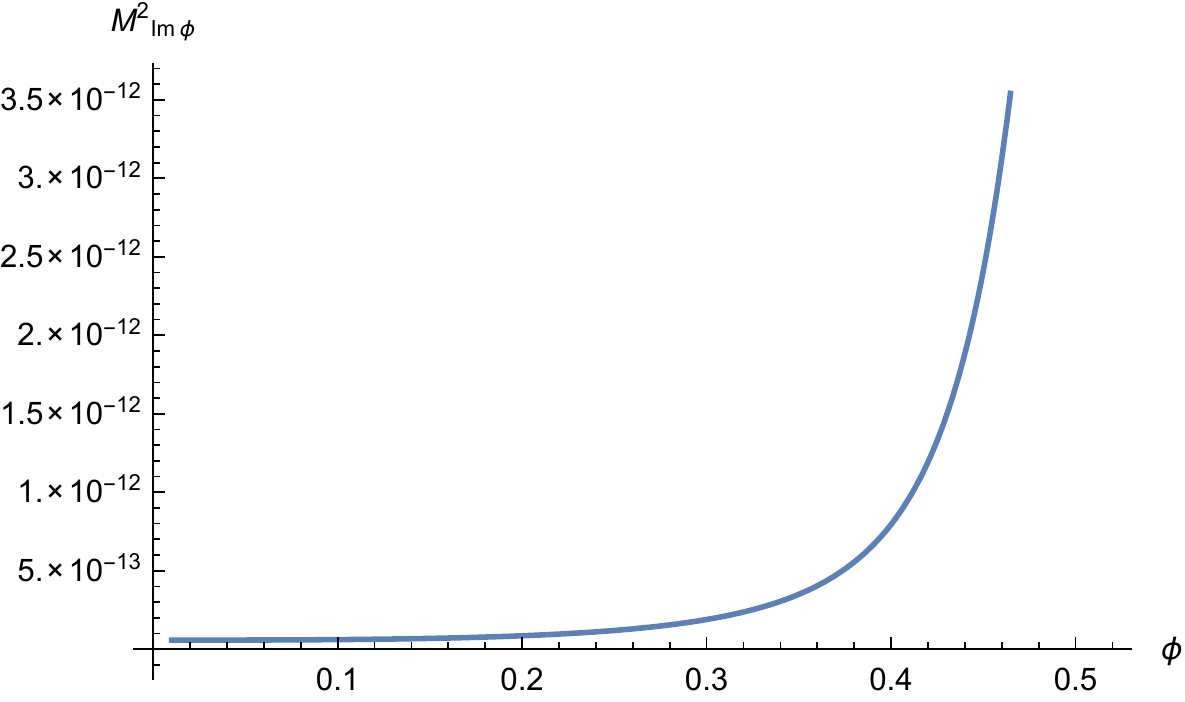}
		\caption{\it Scalar potential along $T$-direction minimum (upper panel) and mass of $\Im(\Phi)$ (lower panel) for the general coupling case. There is a finite positive mass of order Hubble scale at the boundary $\Phi = 0$ and a large positive mass around the vacuum at the end of inflation. (Parameters: $W_0= -0.0004$, $A_+ = 6$, $A_- = 1$, $a = 0.1$, $\alpha = 1$.)}
		\label{fig:minimalgeneralcoupling}
	\end{center}
	\vspace{0cm}
\end{figure}

We now consider a model of the general coupling type as described in Sec.~\ref{Sec.General}. We take $\alpha = 1$ and examine the real directions of the complex scalars first. The superpotential is:
\begin{equation}
W = \Phi^{3}(W_0 + A_+ e^{-aT}) - (W_0 + A_- e^{-aT}) + s_0 \Phi^{\frac{3}{2}}S
\end{equation}
with different parameters $A_\pm$. The $S$ term is included to uplift the AdS vacuum at the end of inflation to Minkowski, thus breaking supersymmetry and still allowing for a gravitino mass significantly below the Hubble scale of inflation. The scalar potential along the $T$-direction minimum is pictured in Fig.~\ref{fig:minimalgeneralcoupling}. 

The waterfall effect ends inflation at $\varphi \simeq 0.43$ in canonical variables or $\Phi \simeq 0.70$ in geometric variables. For $N = 50$, the shift of the effective number of e-folds due to the waterfall is $O(1)$. The shift $\Delta N$ is generally very insensitive to changes of the superpotential parameters. We conclude that the observables of the general coupling model without inflationary profile converge to the universal attractor point \eqref{PredictionsGeneralCoupling} for a large class of parameter choices, for any order unity $\alpha$.

\subsubsection*{Stability of the imaginary directions}
The stability analysis changes slightly due to the introduction of the nilpotent uplifting. This adds a constant contribution to the squared mass of the imaginary $\Phi$ direction. At the boundary, we have:
\begin{align}
M^2_{\Im(\Phi)} =& \frac{a^2 (\alpha -1) A_-^2 e^{-2 a T} \Phi^{-3 \sqrt{\alpha }}}{2^{3 \alpha-1}\ 9 \alpha  T} + \frac{2^{-3 \alpha -2} e^{-2 a T}}{9 \alpha  T^3} \notag\\
&\times\big[-8 a (\alpha +1) A_-^2 T (2 a T+3)\notag\\
&+24 a (\alpha +1) A_+ A_- S+9 \alpha  s_0^2	e^{2 a T}\big]
\end{align}
Just as before, all directions are stable and highly massive in the vacuum at the end of inflation. This vacuum is SUSY AdS without nilpotent uplifting and non-supersymmetric Minkowski with the uplifting. In the $\alpha = 1$ example, the imaginary direction of $\Phi$ is positive throughout the field range of inflation. The mass is of the order of the Hubble scale around $\Phi \rightarrow 0$ and then quickly rises (see Fig.~\ref{fig:minimalgeneralcoupling}). 

If one uses a different uplifting mechanism (e.g. an explicit supersymmetry breaking term induced by anti-branes as consider by KKLT), this raising of the $\Im(\Phi)$ mass at the boundary of moduli space may not occur. There is still a stable region of field space which can support 60 e-folds of inflation, but there is no decoupling of $\Im(\Phi)$ at the boundary. It is possible to solve this using the methods of the previous section. 

\section{General coupling with nilpotent sGoldstino}\label{Sec.GeneralNilpotent}

In the previous sections, we have considered the inflationary sector as given by the single-superfield formulation of $\alpha$-attractors of \cite{AlphaScale}, with no need of additional superfields to assure stabilization. This model has an intricate structure that is necessary to generate a de Sitter solution. We can simplify the superpotential considerably by making more extensive use of uplifting. However, this changes certain qualitative aspects of our previous setup. Our new starting point is the following model with simplified superpotential:
\begin{align}
W = \Phi^{\frac{3\alpha}{2}} W_{mod}(T)\,,
\end{align}
with the standard sum-separable \Kahler potential and the usual KKLT $W_{mod}$. In the modified shift-symmetric \Kahler frame defined by Eq.~\eqref{shift-K}, the overall power of $\Phi$ is gauged away and the superpotential reads simply $W(\Phi, T) = W_{mod}(T)$. 

As a product separable superpotential, the SUSY condition of KKLT carries over exactly and it corresponds to a solution of $\partial_T V = 0$. However, the scalar potential is flat AdS with cosmological constant: 
\begin{equation}
\Lambda = -\frac{2^{-3 \alpha -1} a^2 A^2 e^{-2 a T_0}}{3 T_0}.
\end{equation}
with $T_0$ determined by the SUSY condition \eqref{eq:SUSYKKLT}. We can uplift the scalar potential by means of a nilpotent chiral superfield $S$. This is subject to the constraint $S^2 = 0$. The non-trivial solution to this superfield equation involves writing the scalar part in $S$ as a bilinear in fermions, divided by its auxiliary component. The nilpotent chiral superfield therefore carries no scalar degrees of freedom. We may treat $S$ as a regular superfield, calculate the scalar potential and at the end impose $S = 0$, as advocated in \cite{CosmologyNilpotent} and employed in different cosmological applications (see e.g. \cite{Antoniadis:2014oya,Kallosh:2014via,Dall'Agata:2014oka,Kallosh:2014hxa,deSitterLandscape,Carrasco:2015pla}). To the \Kahler potential we add a canonical term $S\bar{S}$. We choose the following superpotential:
\begin{equation}
W = \Phi^{\frac{3\alpha}{2}}\ W_{mod}(T)\ s_0 S\,,
\end{equation}
which maintains the product separable structure. The cosmological constant becomes
\begin{equation}
\Lambda=\frac{2^{-3 \alpha -1} a^2 A^2 s_0^2 e^{-2 a T_0}}{9 T_0}\,.
\end{equation}
We can generate an inflationary slope by deforming the product separable superpotential. In particular, we can add generic expansions $f(\Phi)$ and $g(\Phi)$ to the $S$-dependent factor in the superpotential such as
\begin{equation}\label{eq:LandscapeModel}
W = \Phi^{\frac{3\alpha}{2}}\left[W_{mod}(T)\right]\left[f(\Phi) +g(\Phi)S\right]\,.
\end{equation}
This breaks the product separable structure between $S$ and $\Phi$, but not between $T$ and the other superfields, so that $D_T W =0$ still defines an extremal trajectory. We recover the model of \cite{deSitterLandscape} along the trajectory defined by $D_T W = 0$.

One can further generalized the model by decomposing the superpotential into all the different types of coupling it contains:
\begin{equation}
W = \Phi^{\frac{3\alpha}{2}}\left[A(\Phi) + B(\Phi)e^{-aT} + C(\Phi)S + D(\Phi)Se^{-aT}\right]\,.
\vspace{0.3cm}
\end{equation}
The latter expression generically leads to $\alpha$-attractor behaviour when we consider the functions $A(\Phi)$, $B(\Phi)$, etc. to be independent generic expansions in $\Phi$. We conclude that making use of a nilpotent sector simplifies the superpotential considerably, as was already noted in \cite{deSitterLandscape}. Specifically the case $\alpha=2/3$ generates a very simple setting with just integer powers of $\Phi$ in $W$ and a simple \Kahler potential such as 
\be
\begin{aligned}
K&=-2\log(\Phi + \bar \Phi) -3 \log(T + \bar T)+ S\bar{S}\,,\\ 
W &= \Phi\left[A(\Phi) + B(\Phi)e^{-aT} + C(\Phi)S + D(\Phi)Se^{-aT}\right]\,.
\end{aligned}
\ee
Note that one can choose the functions $B$, $C$ and $D$ also to be independent of $\Phi$. In addition the three-fields coupling term, parametrized by $D$, can  be set equal to zero. 

In the asymptotic region $\Phi \rightarrow 0$, the location of the $T$-direction minimum $T_0$ is determined entirely by the constant terms in the generic expansion. By choosing these coefficients appropriately, we can generate an inflationary plateau with stabilized volume modulus. Unlike in the previous setup, it is possible to choose profile functions $A(\Phi)$, $B(\Phi)$, etc. such that the scalar potential contains a $T$-direction minimum throughout the entire range $\Phi = (0, \infty)$. This requires tuning one of the constant order coefficients in the expansions if we take polynomial profile functions. The \Kahler modulus minimum $T_0$ as a function of $\Phi$ then smoothly interpolates between its asymptotic limits. Unlike in the previous setup, there is no waterfall effect for a generic choice of profile functions.

The first deviation from the asymptotic $T_0$ at $\Phi = 0$ is of order $\Phi^1$ (when the linear terms in the expansions do not vanish). In geometrically defined variables $\Phi$ and $T$, the field excursions during inflation of $\Phi$ and $T$ are of the same order of magnitude. However, the moduli space geometry around $\Phi = 0$ stretches out the $\Phi$ excursion in the canonical variable $\varphi$, so that the effective single-field description of inflation is justified. In this case, the decoupling of the volume modulus during inflation is entirely due to the boundary point becoming a very long plateau in canonical coordinates. Remember that in the previous setup based on $\alpha$-scale supergravity, there were two effects contributing to the suppression of the \Kahler modulus backreaction: the inverse power ($\Phi^{- \frac{3}{2}\sqrt{\alpha}}$ in shift symmetric \Kahler frame) generated a large mass for $\Re(T)$, and the moduli space geometry stretched out the $\Phi$ excursion in canonical variables.

To end with a concrete example, the choice\footnote{Note that the coefficients not multiplying an exponential in $T$ are much smaller than $B(\Phi) = 1$, as is always necessary in KKLT to stabilize the \Kahler modulus at a large positive value.} $A(\Phi) = -0.0004 -0.0001\ \Phi +0.00005 \  \Phi^2$, $B(\Phi) = 1$, $C(\Phi)= 0.0006825$ and $D(\Phi)=0$ generates an effective scalar potential with a linear fall-off at the boundary of moduli space and a nearly Minkowski minimum at $\Phi \simeq 1$ (i.e. $\varphi \simeq 0$). The mass squared of both imaginary directions $\Im(T)$ and $\Im(\Phi)$ is order Hubble scale or higher throughout the inflationary trajectory. The \Kahler modulus makes a relatively modest field excursion of $\Delta T \simeq 1.7$ from $\Phi = 0$ to the vacuum at $\Phi = 1$, with most of this excursion happening close to $\Phi =1$, at the end of inflation. As the fall-off is linear in geometric variables at the boundary of the moduli space, the predictions are the ones typical of $\alpha$-attractors  Eq.~\eqref{usual-predictions}.

\section{Conclusions}\label{SECconclusions}
 
In this work, we have provided strong evidences for a relative immunity of inflationary $\alpha$-attractors to the backreaction of \Kahler moduli, within the KKLT stabilization scenario. Specifically, we have shown that the effects of a \Kahler modulus $T$ is negligible during the expansion period, which is driven by the real component of the superfield $\Phi$. This phenomenon has been observed in all the three coupling cases analysed in this paper (i.e.~product coupling, general coupling and with nilpotent sGoldstino). 

This stability is intimately connected to the fact that inflation takes place at the boundary of moduli space. In this limit, the coupling of the two sectors produces indeed a number of interesting features. The original KKLT minimum is raised to positive values thanks to the supersymmetry breaking in the $\Phi$-direction. On the other hand, its stability and supersymmetry ($D_{T}W=0$) features remain unaffected. This so-called {$\alpha$-KKLT minimum} becomes a perfect starting point for inflation. Once we switch to the canonical variable for the inflaton field, this boundary point gets indeed stretched to a long dS plateau.

Moving away from the boundary, the inflaton always follows the characteristic exponential fall-off, yielding universal cosmological predictions given by Eq.~\eqref{usual-predictions}. This can always be induced by inserting a profile function $f(\Phi)$ (a generic Taylor expansion) into $W$, analogously to what happens to the original $\alpha$-attractor models \cite{SuperconformalAttractors,DoubleAttractors,Kallosh:2015lwa,AlphaScale,deSitterLandscape,Carrasco:2015pla}. More interestingly, we have shown that, in the case of general couplings (analysed in Sec.~\ref{Sec.General}), the exponential deviation from a positive plateau simply becomes a genuine and natural consequence of the mutual backreaction between $T$ and $\Phi$. In the latter case, the observational prediction are universal and restricted to Eq. \eqref{PredictionsGeneralCoupling}.

Approaching the end of inflation, the interplay between the modulus $T$ and $\Phi$ does become important. It produces a waterfall effects which ends inflation and leaves all the scalars in a phenomenologically suitable vacuum. This vacuum is supersymmetric, in absence of any uplifting mechanism to de Sitter. Remarkably, this means that there is generically no connection between the gravitino mass in the vacuum and the Hubble scale of inflation. Although some proposals have pointed out how to decouple these physical scales \cite{Kallosh2004}, our results suggest a new approach to solving this problem.

The above findings represent a novelty in the landscape of previous studies about the interaction between moduli and inflation. Especially in the case of large-field scenarios, the claims have been often negative: the backreaction of the \Kahler modulus was destabilizing the original inflationary dynamics. In the most optimistic scenario, a certain amount of fine-tuning was required in order to generate the minimum amount of e-folds of exponential expansion, although with some modification of the original inflationary predictions. In \cite{Challenges} this effect was dubbed ``flattening'' as it generically lowered the value of the tensor-to-scalar ratio with respect to the original $\phi^2$- predictions. 

The present study appears to be free of such problems: the backreaction of the moduli does not destabilize the inflationary trajectory. Instead, the additional sector induces an inflationary profile in the case of general couplings, and moreover it offers a universal supersymmetric minimum to the scalars after inflation. Moreover, strict bounds between the value of the gravitino mass and the inflationary scale were always representing a threat to model-builders.

A number of aspects deserve further study. On the phenomenological side, these include a detailed investigation of the choice of the inflationary profile. We have provided a proof of principle that one can either introduce a tailor-made Minkowski minimum along the asymptotic-KKLT branch, or end up in the universal Minkowski minimum along the other branch. It remains to be seen what is generic, and how stable various choices are. In contrast, on the string theory side, it remains a challenge of embedding $\alpha$-attractors in a concrete scenario. Despite some approximate realizations in specific contexts (see e.g. \cite{Cicoli:2008gp,Broy:2015zba} for fibred Calabi-Yau geometries), one would like to identify the natural mechanism underlying the attractor nature of these models, once the appropriate inflaton modulus sector has been recognized. In this respect, the present study provides a useful guideline to determine the generic structure which always preserves the asymptotic inflationary plateau at the boundary of moduli space. Whereas the hyperbolic \Kahler geometry of the inflaton plays again a crucial role, the  coupling patterns here discussed leave a certain freedom for the superpotential. The general coupling set-up \eqref{eq:GeneralCouplingSuper} seems most promising, as it essentially consists of two copies of no-scale KKLT. It requires no additional inflationary profile and has universal observational predictions \eqref{PredictionsGeneralCoupling}.

\section*{Acknowledgments}
We would like to thank Renata Kallosh, Andrei Linde, Alexander Westphal, Clemens Wieck and Timm Wrase for very stimulating discussions and useful comments on a draft of this paper. MS gratefully acknowledges financial support by the collaborative research center SFB 676 and by `The Foundation Blanceflor Boncompagni Ludovisi, n\'ee Bildt'. PW acknowledges financial support by the Dutch Foundation for Fundamental Research on Matter (FOM).

\bibliography{results}
\bibliographystyle{utphys}

\end{document}